\documentclass[11pt,letterpaper]{article}
\pdfoutput=1
\usepackage{latexsym}
\usepackage{amsthm}
\usepackage{amsfonts}
\usepackage{authblk}
\usepackage{lipsum}
\usepackage{amssymb}
\usepackage{paralist}
\usepackage{amsmath}
\usepackage{xspace}
\usepackage{mathrsfs}
\usepackage[letterpaper,margin=3cm]{geometry}
\usepackage{changepage}
\usepackage{appendix}
\usepackage[dvipsnames,usenames]{xcolor}
\usepackage{graphicx}
\usepackage[boxed,figure,vlined]{algorithm2e}
\usepackage{enumerate}
\usepackage{tabularx}
\usepackage{subcaption}
\usepackage{caption}
\usepackage{color}
\usepackage{hyperref}
\usepackage[capitalize, nameinlink]{cleveref}

\definecolor{darkred}{rgb}{0.75,0,0}

\definecolor{darkgreen}{rgb}{0,0.5,0}
\hypersetup{
    unicode=false,          
    colorlinks=true,        
    linkcolor=red,          
    citecolor=darkgreen,        
    filecolor=magenta,      
    urlcolor=cyan           
}
\crefname{theorem}{Theorem}{Theorems}
\crefname{lemma}{Lemma}{Lemmas}

\providecommand{\keywords}[1]{\textbf{Keywords:} #1}

\newcommand{\opt}{\ensuremath{\mathcal{O}}}
\newcommand{\on}{\ensuremath{\mathcal{ON}}}
\newcommand{\off}{\ensuremath{\mathcal{OF}}}
\newcommand{\givenalg}{\ensuremath{\mathcal{A}}}
\newcommand{\DGA}{DGA}
\newcommand{\dga}{\ensuremath{\mathcal{G}}}
\newcommand{\algsol}{\ensuremath{\mathcal{F}}}
\newcommand{\cost}{\ensuremath{\mathit{cost}}}
\newcommand{\metric}{\ensuremath{\mathcal{M}}}
\newcommand{\fac}{\ensuremath{\mathcal{S}}}
\newcommand{\dem}{\ensuremath{\mathcal{J}}}
\newcommand{\assign}{\ensuremath{\mathcal{I}}}
\newcommand{\service}{\sigma}
\newcommand{\Service}{S}
\newcommand{\ServiceLB}[1]{\overline{\Service}_{#1}}
\newcommand{\etaLB}[1]{\overline{\eta}_{#1}}
\newcommand{\ndem}{r}
\newcommand{\nfac}{f}
\newcommand{\move}{M}
\newcommand{\movement}{m}
\newcommand{\ptime}{\theta}
\newcommand{\mtime}{\tau}
\newcommand{\Dest}{V^{\mathit{dst}}}
\newcommand{\src}{v^{\mathit{src}}}
\newcommand{\dest}{v^{\mathit{dst}}}
\newcommand{\mtimealg}{\tau^{\dga}}
\newcommand{\srcalg}{v^{\mathit{src},\dga}}
\newcommand{\destalg}{v^{\mathit{dst},\dga}}
\newcommand{\fpar}{\alpha}
\newcommand{\spar}{\beta}
\newcommand{\ch}{d}

\newcommand{\floor}[1]{\left\lfloor #1 \right\rfloor}
\newcommand{\RR}{\ensuremath{\mathbb{R}}}
\newcommand{\eps}{\varepsilon}
\newcommand{\vect}[1]{{\boldsymbol{#1}}}
\newcommand{\E}{\ensuremath{\mathbb{E}}}

\renewcommand{\include}{\input}

\newcommand{\para}[1]{\medskip\noindent\textbf{#1}}

\newtheorem{theorem}{Theorem}[section]
\newtheorem{lemma}[theorem]{Lemma}
\newtheorem{claim}[theorem]{Claim}
\newtheorem{corollary}[theorem]{Corollary}
\newtheorem{definition}{Definition}[section]
\newtheorem{observation}[theorem]{Observation}
\newtheorem{remark}{Remark}[section]

\title{Toward Online Mobile Facility Location on General Metrics\thanks{This work is based on material that appears \cite{ghodselahi2018serving}. A preliminary version of this paper was presented in the Proceedings of the 26th International Symposium on Algorithms and Computation (ISAAC 2015) \cite{ghods2015}. Additionally, the comprehensive version has been published in the Journal of Theory of Computing Systems (TOCS) \cite{ghodselahi2023toward}.}}

\author{Abdolhamid Ghodselahi\thanks{University of Passau, 94032 Passau, Germany, \texttt{abdolhamid.ghodselahi@uni-passau.de}} \quad \quad
Fabian Kuhn\thanks{University of Freiburg, 79110 Freiburg, Germany, \texttt{kuhn@cs.uni-freiburg.de}}}
        
\date{}

\begin{document}

\maketitle              

\begin{abstract}
We introduce an online variant of mobile facility location (MFL) (introduced by Demaine et al. [SODA' 07]). We call this new problem \emph{online mobile facility location} (OMFL).
In the OMFL problem, initially, we are given a set of $k$ mobile facilities with their starting locations. 
One by one, requests are added.
After each request arrives, one can make some changes to the facility locations before the subsequent request arrives. Each request is always assigned to the nearest facility. The cost of this assignment is the distance from the request to the facility.
The objective is to minimize the total cost, which consists of the relocation cost of facilities and the distance cost of requests to their nearest facilities.
\\
We provide a lower bound for the OMFL problem that even holds on \emph{uniform metrics}.
A natural approach to solve the OMFL problem for general metric spaces is to utilize \emph{hierarchically well-separated trees} (HSTs) and directly solve the OMFL problem on HSTs.
In this paper, we provide the first step in this direction by solving a \emph{generalized variant} of the OMFL problem on uniform metrics that we call G-OMFL.
We devise a simple deterministic online algorithm and provide a tight analysis for the algorithm.
The second step remains an open question.
Inspired by the $k$-server problem, we introduce a new variant of the OMFL problem that focuses solely on minimizing movement cost. We refer to this variant as M-OMFL. Additionally, we provide a lower bound for M-OMFL that is applicable even on uniform metrics.
\end{abstract}

\section{Introduction}
\label{sec:intro}
On-demand mobile distribution hubs are emerging as a popular solution for businesses looking to meet growing demand and provide efficient, long-term access to warehouse space for their customers. These mobile facilities can be relocated to different locations within a region, based on market needs, and can be used for various purposes such as delivering goods, providing services, and improving network coverage. To reduce different costs, applications may need to move facilities closer to the end user. This means that facilities should always stay within a reasonable distance from customers who have ongoing service demands. However, relocating facilities also involves a cost. An efficient solution could dynamically adjust the locations of these mobile facilities in response to new customers, ensuring they are always within a reasonable distance of where they are needed. This results in a trade-off between two types of costs: the assignment cost, which is the sum of distances between customers and their nearest facilities at the current time, and the total movement cost, which is the total cost of relocating facilities between locations so far. The goal is to minimize the total cost of serving customers that arrive over time and request long-term service, which is the sum of these two costs.

In this paper, we introduce the online mobile facility location (OMFL) problem. We present two applications from different domains. The first one is related to the distribution of goods from mobile warehouses to retail stores. Suppose a company has a fixed number of mobile warehouses, initially located at various points within a region. The company serves this region, where the distances between points represent the cost of shipping goods or relocating warehouses. As demand for the company's products grows, new retail stores are built, resulting in new requests. These requests represent new retail stores that need to be served by the company. In this scenario, we do not consider the individual demands of each retail store. Instead, we assume that these stores typically have demands for goods that need to be shipped from mobile warehouses.

In the second application, we consider a telecommunications company that provides mobile broadband services to households within a region using movable internet providers. The company has a fixed number of these providers, which are initially positioned at various locations throughout the region. The distances between these points represent the signal strength or coverage between them. As the population in the region gradually grows and new households are established or connected, there is an increase in demand for the company's services, resulting in new requests for continuous and long-term mobile broadband coverage.

\para{Paper Outline} In the rest of this section, we introduce the OMFL problem and its two variants, G-OMFL and M-OMFL. Each subsection defines the corresponding problem, provides results, and reviews related work. In Section~\ref{sec:G-OMFL}, we analyze G-OMFL and provide an upper bound. Section~\ref{sec:OMFL} presents our lower bound analysis for OMFL, showing that our analysis for G-OMFL is almost tight since OMFL is a special case of G-OMFL. Section~\ref{sec:M-OMFL} offers our lower bound analysis for M-OMFL. Finally, Section~\ref{sec:conclusion} concludes the paper and poses two open questions. Appendix~\ref{sec:app_hst} defines HSTs and discusses their use as a tool for solving online problems.

\subsection{Online Mobile Facility Location (OMFL)}
\label{sec:intro_omfl}
The mobile facility location (MFL) problem was first introduced by Demaine et al. \cite{demaine2007minimizing} and later studied by Friggstad and Salavatipour \cite{friggstad2008minimizing}. Many MFL scenarios, such as the two applications presented above, are online, where requests arrive one at a time and require long-term service, and the end of the request sequence is unknown. In MFL, each facility and request has a starting location in a metric space, and the goal is to find a destination point for each such that every request is assigned to a point that is the destination of some facility. The objective can be either to minimize the total or maximum distance between any pair of facilities and requests and their destinations. In OMFL, facilities can move more than once as new requests arrive online, while in MFL, facilities move only once offline. However, the assignment cost in both problems is based on the final configuration of facilities. In the online setting, the future requests are unknown, so any facility configuration at a given time could be the final one. The intermediate movements of facilities could be interpreted as the ``learning cost'' that an online algorithm must pay compared to an offline algorithm that solves MFL to ensure that facilities are well positioned at any time. Therefore, OMFL can be seen as an online adaptation of MFL, where the online algorithm must balance the trade-off between moving facilities closer to requests and minimizing relocation costs.

\subsubsection{Problem Defintion}
\label{sec:intro_omfl_problem}

We define an instance of the OMFL problem as follows. We are given an arbitrary finite metric space $\metric=(V,d)$ where $\lvert V \rvert = n$ and a non-negative, symmetric distance function $d:V \times V \rightarrow \RR_{+}$, which satisfies the triangle inequality. A set of requests are issued at points one at a time. There is a set of $k$ mobile facilities. We define a configuration (at the time $t=1,2,\ldots$) to be a function $\fac(t): [k]\rightarrow V$ that specifies which of the facilities $\{1,\ldots,k\}$ is located at which point at time $t$, i.e., $\fac_{j}(t)$ denotes the point that hosts facility $j$ at time $t$. At each time $t=1,2,3,\dots$, there is exactly one request placed, and $\dem(t)$ denotes the point at which the single request at time $t$ is placed.

Formally, ``serving all requests" at time $t$ is a function $\assign: [t] \rightarrow [k]$,
which assigns any request $i \in [t]$ to one of the $k$ facilities\footnotemark[1]\footnotetext[1]{We remark that for two integers $a\leq b$, $[a,b]:=\{a,\dots,b\}$ and for an integer $a\geq 1$, we use $[a]$ as a short form to denote $[a]:=[1,a]$.}. We assume there is no bound on the
number of requests that can be assigned to the same facility. If request $i$ is assigned to facility
$j$ at time $t$ then this induces the cost $d\left(\dem(i),\fac_j(t)\right)$, denoted as {\em assignment cost}
of request $i$ at time $t$.
The {\em overall assignment cost} of a configuration $F \subset V$ at time $t$
denoted by $\Service_t(F)$ is the sum of the assignment costs of all requests at this time, i.e.,
$\sum_{i=1}^t d\left(\dem(i),\fac_{\assign_i(t)}(t)\right)$.
In order to keep the overall assignment cost
small, an algorithm can move the facilities between the points, which implies some additional
cost, called the {\em movement cost}. The movement cost equals the distance between the
points, i.e., if facility $j$ is at point $v$ at time $t$ and at point $v'$ at time $t+1$, the cost of
movement of facility $j$ at time $t+1$ equals the distance $d(v,v')$.

\para{Feasible Solution} At any given time $t$, any facility configuration at time $t$ is considered a feasible solution. This means that even if an algorithm does not perform any actions can output the initial facility configuration as a feasible solution. However, this solution may not necessarily be efficient, as it may result in a high assignment cost.

The total cost of any algorithm \givenalg\ at time $t$ is the sum of the total movement cost by
time $t$ and the overall assignment cost at time $t$ and defined as follows
\begin{equation}\label{eq:objective_OMFL}
	\cost_{t}^{\givenalg}:=\Service^{\givenalg}_t+\move^{\givenalg}_t
\end{equation}
where
\[
	\Service^{\givenalg}_t:=\sum\limits_{i=1}^t d\left(\dem(i),\fac^{\givenalg}_{\assign^{\givenalg}_i(t)}(t)\right)
\]
and 
\[
	\move^{\givenalg}_t:=\sum\limits_{i=1}^t\sum\limits_{j=1}^{k} d\left(\fac^{\givenalg}_j(i-1),\fac^{\givenalg}_j(i)\right).
\]

\begin{remark}\label{re:competitive_ratio}
The \textbf{competitive ratio} for any online algorithm at any given time $t \geq 1$ is defined as the total cost of the online algorithm at time $t$ divided by the cost of the optimal MFL solution at time $t$. It is important to note that an MFL algorithm can wait until time $t$ and perform all necessary facility movements at that time. However, the online algorithm does not know the value of $t$ in advance. As a result, if the online algorithm aims to guarantee a competitive ratio $\gamma$ against the optimal MFL solution at any given time $t$, it must guarantee $\gamma$ at any time $t' \in [t]$ and may relocate facilities at any time $t'$ since it does not know the value of $t$ in advance.
\end{remark}

\begin{definition}[Optimal Assignment Cost]
\label{def:optimal_assign}
The optimal assignment cost at any time $t$, denoted by $\Service_t^*$, is the lowest total assignment cost that can be achieved by an ``optimal configuration'' of facilities at time $t$.
Mathematically,
\[
\Service_t^* := \min_{F} \Service_t(F),
\]
where $\Service_t(F)$ is the sum of the assignment costs of all requests at time $t$ given the configuration $F$ of facilities at time $t$. Further, $F^* \in \arg\min\limits_{F}\Service_t(F)$ is an optimal configuration.
\end{definition}

\begin{remark}\label{re:optimal_assign}
Due to the trade-off in the cost function~\eqref{eq:objective_OMFL}, an optimal offline algorithm \opt\ may not move facilities to the optimal configuration. This means that the optimal offline algorithm may lose more in movement cost than it gains in assignment cost if it moves its facilities to the configuration that minimizes the total assignment cost. Therefore, the total assignment cost of an optimal offline algorithm at any time $t$, i.e., $\Service_t^{\opt}$ may not equal the optimal assignment cost at $t$, i.e., $\Service_t^*$. 
\end{remark}

\subsubsection{Our Results}
\label{sec:intro_omfl_result}
We provide a lower bound on the achievable competitive ratio. Moreover, the lower bound even holds for the OMFL problem on uniform metric spaces when all pairwise distances are equal.
The lower bound theorem essentially says that if any online algorithm \on\ wants to guarantee
a small multiplicative competitive ratio, \on\ has to tolerate a
relatively large additive term. Basically, if \on\ wants to keep the
additive term within a bound of $\beta$, the multiplicative
competitive ratio becomes at least $1+\Omega(k/\beta)$.

\begin{theorem}\label{thm:LB_OMFL}
  Let \opt\ denote an optimal offline algorithm.
  Consider any deterministic online algorithm
  \on. Further, assume that \on\ guarantees that at all
  times $t>0$, the additive difference between the assignment cost of
  $\on$ and the optimal assignment cost
  at time $t$ is less than $\beta$. Then, there
  exist an execution and a time $t_0 > 0$ such that the total cost of
  \on\ can be lower bounded as follows.
  \begin{itemize}
  \item If $\beta=O(k/\eps)$ for any $\eps>0$, it
    holds that
    \[
      	\cost_{t_0}^\on \geq
        (1+\eps ) \cdot 
        \cost_{t_0}^\opt+ \Omega(k  \log k - \eps k).
    \]
  \item If $\beta = O\left(\frac{k \cdot \log
        k}{\log\log k}\right)$, then for every $\eps \geq \frac{\log \log
    k}{\log^{1-\delta}k}$ and any constant $0 < \delta \leq 1$, we obtain
    \[
    \cost_{t_0}^\on \geq
     (1+\eps ) \cdot 
        \cost_{t_0}^\opt + \Omega\left(\frac{k \cdot \log k}{\log\log
            k}-\eps k\right).
    \]
  \end{itemize}
\end{theorem}

\subsubsection{Related Work}
\label{sec:intro_omfl_related}
MFL is a generalization of the $k$-facility location problem \cite{devanur2005price}, which itself generalizes the facility location and $k$-median problems \cite{arya2001local,jain2002new,drezner2004facility}. \cite{demaine2007minimizing} provided a 2-approximation for the maximum objective, which is optimal unless $P=NP$. \cite{friggstad2008minimizing} provided an 8-approximation for the total objective.

Consider the following abstract online problem: The problem is defined on a graph or metric space with resources that can be either fixed (potentially any point of the metric can be opened as a resource) or mobile (typically $k$ resource points are initially given). Requests are revealed one by one and must be served by a resource. A request may need to be served by moving a mobile resource to the requested point. Alternatively, a model may relax this consideration and allow a request to be served remotely by either a fixed or mobile resource. The offline setting for resource allocation problems does not clearly capture the difference between real-world applications where requests need long-term service or one-time service at the time of arrival. However, the online setting can define this difference more clearly. Requests may need either long-term service or to be served only at the time they arrive. The assignment cost of a request, $r_i$, that arrives at time $i$ for one-time service is equal to the distance from $r_i$ to $f_i(i)$, where $f_i(t)$ represents the nearest facility to $r_i$ at any time $t \geq i$. For long-term service, there are different cases. Consider the following two cases: 1) The ``fixed long-term" case, where the reassignment is not an option, the assignment cost of $r_i$ is the distance from $r_i$ to the point that hosts $f_i(i)$ at any time $t \geq i$. 2) The ``dynamic long-term" case, where the reassignment is possible, the assignment cost of $r_i$ equals the distance from $r_i$ to $f_i(t)$ at any time $t \geq i$. The movement cost is the total distance resources travel. In some models, moving a resource over some distance is more expensive than serving a request over the same distance, that we call it ``expensive move''. By varying these criteria, several classic online problems are defined. Some aim to minimize only the total movement cost, while others involve a trade-off between the assignment cost and either the total opening cost or the total movement cost. The goal is to find an optimal balance between these costs.

OMFL is an online problem that falls under the umbrella of this abstract problem, along with other classic online problems such as page migration introduced by Black and Sleator \cite{black1989competitive}, online facility location (OFL) introduced by Meyerson \cite{meyerson2001online}, and $k$-server introduced by Manasse et al. \cite{manasse1988server}. For uniform metric spaces, the $k$-server problem is equivalent to the paging problem \cite{sleator1985amortized}. Table~\ref{tab:my-table} summarizes the similarities and differences between OMFL and these classic online problems, as well as some other related work.

\begin{table}[htbp]
\centering
\resizebox{\textwidth}{!}{%
\begin{tabular}{p{3.9cm}||p{0.8cm}|p{1.5cm}|p{2cm}|p{2.5cm}}
\hline
\multicolumn{1}{c}{\textbf{Problem}} & \multicolumn{1}{c}{\textbf{Type}} & \multicolumn{1}{c}{\textbf{Location}} & \multicolumn{1}{c}{\textbf{Duration}} & \multicolumn{1}{c}{\textbf{Cost}}\\
\hline
\hline
OMFL & mobile & remote & dynamic long-term & move+assign \\
\hline
Online $k$-Facility Reallocation \cite{de2022facility,fotakis2021reallocating} & mobile & remote & one-time & move+assign \\
\hline
$k$-Page Migration & mobile & remote & one-time & expensive move+assign \\
\hline
$k$-Server & mobile & requested point & one-time & move \\
\hline
OFL & fixed & remote & one-time & open+assign\\
\hline
OFL with Mobile Facilities \cite{Feldkord2022online}& mobile & remote & one-time / fixed long-term & expensive move+open+assign\\
\hline
\end{tabular}%
}
\caption{A comparison of OMFL and related work in terms of resource type, service location, service duration, and cost function.}
\label{tab:my-table}
\end{table}

In the online version of the $k$-facility reallocation problem, we have more than one customer points that request service at each time step, instead of one request at a time as in the abstract online problem. Fotakis et al. \cite{fotakis2021reallocating} present a constant-competitive algorithm for this setting when $k = 2$ on the lines.

Almost all work have investigated the page migration problem for only a single page (i.e., $k=1$). A recent paper introduces a variant of page migration in the Euclidean space (of arbitrary dimension) where they limit the allowed distance the facility can move in a time step \cite{feldkord2019mobile}. \cite{feldkord2019managing} extends \cite{feldkord2019mobile} for $k$ mobile facilities. The best deterministic and randomized algorithms for page migration problem have competitive ratios of $4$ \cite{bienkowski2019dynamic} and $3$ \cite{westbrook1994randomized}, respectively. There are methods to transform any $k$-server algorithm into a deterministic and randomized $k$-page migration algorithm with competitive ratios of $O(c^2)$ and $O(c)$, respectively, where $c$ is the competitive ratio of the $k$-server algorithm \cite{bartal2001page}. Results on page migration are elaborated in detail in \cite{bienkowski2012migrating}. 

Fotakis \cite{fotakis2008competitive} shows that the competitive ratio for OFL is $\Theta(\log m / \log \log m)$ where $m$ is the total number of requests. A variant of OFL is the incremental facility location problem \cite{fotakis2004incremental,fotakis2011online} where it is possible to merge clusters. A relaxed variant of incremental facility location that investigates different models is presented where an online algorithm may make corrections to the positions of its open facilities \cite{diveki2011online,Feldkord2022online}. While the OMFL problem is an online variant of MFL, the models proposed by \cite{diveki2011online,Feldkord2022online} are mobile variants of OFL. As such, there are key differences between the OMFL and these models. In contrast to the OMFL problem, \cite{diveki2011online} does not charge any cost for moving a facility. This is a major limitation of their model. They present an algorithm which is $2$-competitive. Afterward, a model was proposed by Feldkord et al. \cite{Feldkord2022online} where an online algorithm may move its open facilities, but instead of this being free, it incurs an ``expensive movement'' cost. Further, they consider two models where the movements of the open facilities can be arbitrarily or limited to some constant in each time step. They achieve the same competitiveness on the line for both one-time and fixed long-term services. In the scenario where movement is unrestricted, the expected competitive ratio achieved is $O(\log D / \log \log D)$, where $D > 1$ represents the multiplicative factor that increases the cost of movement relative to the distance traveled. In the case of limiting the movement of an open facility to some constant $\delta$ in each time step, they achieve an expected competitive ratio that depends on $\delta$ and the cost of opening a new facility. They showed that their results are asymptotically tight on the line. For the one-time service, they extend their result to the Euclidean space of arbitrary dimension, where they achieve a competitive ratio with an additional additive term $O(\sqrt{k})$, where $k$ is the number of facilities in the optimal solution.

Sections~\ref{sec:intro_g-omfl_related} and ~\ref{sec:intro_m-omfl_related} review the similarities and differences between OMFL variants, paging, and the $k$-server problems. M-OMFL is more closely related to paging when it is studied on uniform metrics, while some works on the \emph{randomized $k$-server conjecture} are more closely related to G-OMFL.

\subsection{G-OMFL}
\label{sec:intro_g-omfl}
According to Theorem~\ref{thm:LB_OMFL}, if a company aims to keep its facilities close to an optimal configuration in response to requests at any given time, it is inevitable to have a multiplicative competitive ratio of $\Omega(k)$. This motivates us to explore the role of randomization in the OMFL problem. Following the approach of \cite{cote2008randomized,bansal2011polylogarithmic}, who used randomization to achieve a poly-logarithmic competitive ratio for the $k$-server problem, we also consider solving OMFL on HSTs. In \cite{bansal2011polylogarithmic}, it is shown that by exploiting the randomized low-stretch hierarchical tree decomposition of \cite{fakcharoenphol2003tight}, it is possible to obtain a poly-logarithmic competitive ratio for the $k$-server problem. Utilizing HSTs as a tool in solving some online problems is a common approach \cite{cote2008randomized,bansal2011polylogarithmic,ghods2017arrow,azar2017online} (see Appendix~\ref{sec:app_hst} for more details about HSTs and the approach). HSTs are a useful tool for solving some online problems, as they have a simple structure that allows reducing the problem on an HST to a more general problem on a uniform metric \cite{cote2008randomized,bansal2011polylogarithmic}. This leads us to define a generalized variant of OMFL on uniform metrics as the first step toward solving OMFL on HSTs and general metrics.

\subsubsection{Problem Defintion}
\label{sec:intro_g-omfl_problem}
We adapt the OMFL problem statement to accommodate the G-OMFL on uniform metrics and introduce new notations accordingly. 
We are given a set $V$ of $n$ nodes and there is a set of $k$
facilities. Further, a set of requests arrive one at
a time. We assume that at time $t\geq 1$, request $t$ arrives at node $v(t)\in
V$. For a node $v\in V$, let $\ndem_{v,t}$ be the number of requests
at node $v$ after $t$ requests have arrived, i.e.,
\[
	\ndem_{v,t}:=\lvert \{i\leq t: v(i)=v\} \rvert.
\]
In order to keep
the total service cost small, an online algorithm can move the facilities
between the nodes (if necessary, for answering one new request, we
allow an algorithm to also move more than one facility). We define a \emph{configuration} of facilities
by integers $f_v\in\mathbb{N}_0$ for each $v\in V$ such that $\sum_{v
  \in V} f_v=k$. We describe such a configuration by a set of pairs as
\[
	F := \{(v,f_{v}) : v \in V\}.
\]
The initial configuration is denoted
by $F_0$.

\para{Feasible Solution} The feasible solution for G-OMFL is the same as for OMFL, as defined in Section~\ref{sec:intro_omfl_problem}. This means that any facility configuration at any time step is a feasible solution.

For any online algorithm \on, we denote
the sequence of facility configurations until time $t$ by $\algsol^\on_t:=\{F^\on(i): i \in [0,t]\}$, where $F^\on(t)$ is the configuration after reacting to the arrival of request $t$ and where $F^\on(0)=F_0$.

\para{Service Cost} We implicitly assume that if a node $v$ has
some facilities, all requests at $v$ are served by these facilities.
Depending on the number of facilities and the
number of requests at a node $v\in V$, an algorithm has to pay some
service cost to serve the requests located at $v$. This service
cost of node $v$ is defined by a \emph{service cost function}
$\service_v$ such that $\service_v(x,y)\geq 0$ is the cost for serving
$y$ requests if there are $x$ facilities at node $v$. For convenience,
for $t\geq 1$, we also define
$\service_{v,t}(x):=\service_v(x,\ndem_{v,t})$ to be the service cost
with $x$ facilities at node $v$ at time $t$.  For some configuration
$F$, we denote the total service cost at time
$t$ by
\[
	\Service_t(F) := \sum_{v\in V} \service_{v,t}(f_v) = \sum_{v\in V} \service_v(f_v, \ndem_{v,t}).
\]

The service cost of any algorithm \givenalg\ at time $t$ is denoted by
$\Service_t^\givenalg := \Service_t(F^\givenalg(t))$. Let $\Service_t(F^*)$ equal $\Service_t^*$ for any optimal configuration $F^*$ at time $t$, i.e., $\Service_t(F^*)=\Service_t^*$ (see Definition~\ref{def:optimal_assign} and Remark~\ref{re:optimal_assign} for $\Service_t^*$). 

\para{Movement Cost} We define the movement cost $\move_t^\givenalg$ of
given algorithm \givenalg \ to be the total number of facility movements by time
$t$. Generally, for two configurations, $F=\{(v,f_v): v\in V\}$
and $F'=\{(v,f_v'): v\in V\}$, we define the distance $\ch(F,F')$
between the two configurations as follows:
\begin{equation}\label{eq:serverdistance_GUOMFL}
	\ch(F,F'):=\sum_{v \in V} \max\{0,f_v-f_v'\} =
        \frac{1}{2}\cdot\sum_{v\in V} \lvert f_v - f_v' \rvert.
\end{equation}
The distance $\ch(F,F')$ is equal to the number of movements that are
needed to get from configuration $F$ to configuration $F'$ (or vice
versa). Based on the definition of $\ch$ given by~\eqref{eq:serverdistance_GUOMFL}, we can express the
movement cost of an online algorithm \on\ with the sequence of facility configurations
$\algsol_t^\on=\{F^\on(i): i \in [0,t]\}$ as
\[
\move_t^\on = \sum_{i=1}^{t} \ch\left(F^\on(i-1),F^\on(i)\right).
\]

Regarding the movement cost of an offline algorithm for G-OMFL at any given time $t$, it is similar to the MFL solution in that an offline algorithm for G-OMFL only performs movements once, since it knows the request sequence until time $t$ in advance from the beginning. Therefore, at any given time $t$, the movement cost of any optimal offline algorithm, denoted as \opt, is $\move_t^\opt =  \ch\left(F_0,F^\opt(t)\right)$.

The total cost of any algorithm \givenalg\ is
\[
	\cost_t^\givenalg:=\Service_t^\givenalg + \move_t^\givenalg.
\]

\para{Service Cost Function Properties}
The service cost function $\service$ has to satisfy a number of natural
properties. First of all, for every $v\in V$, $\service_v(x,y)$ has to
be monotonically decreasing in the number of facilities $x$ that are
placed at node $v$ and monotonically increasing in the number of
requests $y$ at $v$.
\begin{eqnarray}
  \forall v\in V\, \forall x,y\in \mathbb{N}_0 & : & 
  \service_v(x,y) \geq \service_v(x+1,y)\label{eq:monotonicfac_GUOMFL}\\
  \forall v\in V\, \forall x,y\in \mathbb{N}_0 & : & 
  \service_v(x,y) \leq \service_v(x,y+1)\label{eq:monotonicdem_GUOMFL}
\end{eqnarray}
These two properties imply that adding more facilities at a node will reduce the total service cost, while adding more requests at a node will increase the total service cost. Naturally, this is because more facilities can serve a fixed number of requests with lower cost, while more requests are served with the same number of facilities with higher cost.

Further, the effect of adding additional facilities to a node $v$
should become smaller with the number of facilities (convex property in $x$)
and it should not decrease if the number of requests gets larger. Therefore, for all
$v\in V$ and all $x,y\in \mathbb{N}_0$, we have
\begin{eqnarray}
  \service_v(x,y)-\service_v(x+1,y)  & \geq &
  \service_v(x+1,y)-\service_v(x+2,y)\label{eq:monotonicdifffac_GUOMFL}\\
  \service_v(x,y)-\service_v(x+1,y) & \leq &
  \service_v(x,y+1)-\service_v(x+1,y+1)\label{eq:monotonicdiffdem_GUOMFL}
\end{eqnarray}
The third property means that if one adds more facilities at node $v$, the total service cost at that node will decrease, but the decrease will be less and less as one adds more facilities. This is because there are diminishing returns to adding more facilities: the first few facilities will have a big impact on reducing the service cost, but after a certain point, the impact will become smaller and smaller. The fourth property means that the difference in assignemnt cost between adding one facility with more requests should be greater than or equal to the difference in cost between adding one facility with fewer requests. These properties ensure that the service cost function $\service$ is well-behaved. 

\begin{remark}\label{rem:especial_case}
We note that the OMFL problem on uniform metric spaces is a special case of G-OMFL, where the cost is
either $0$ (if there is a facility at the node) or it is equal to the number of requests at the node.
\end{remark}

\subsubsection{Our Results}
\label{sec:intro_g-omfl_result}

We devise a simple, deterministic online algorithm, called \emph{dynamic greedy allocation (\DGA)}, denote by \dga with the following properties.
For two parameters $\alpha\geq 1$ and $\beta\geq 0$, \DGA\ guarantees that at
all times $t\geq0$, $\Service_t^\dga < \alpha\Service_t^*+\beta$.
\DGA\ achieves this while keeping the total movement cost small.
In particular, our algorithm a) only moves when it needs to move
because the configuration is not feasible anymore and b) always moves
a facility which improves the service cost as much as possible.
We show that the total number of movements up to a time $t$ of our online
greedy algorithm can be upper bounded as a function of the optimal
service cost $\Service_t^*$ at time $t$.
Most significantly, we show that at the cost of an additive term which is roughly linear in $k$, it is possible to achieve a
competitive ratio of $(1+\eps)$ for every constant $\eps>0$.
This result almost matches the lower bound of Theorem~\ref{thm:LB_OMFL}.

\begin{remark}\label{rem:lowerbound}
Note that the lower bound of Theorem~\ref{thm:LB_OMFL}
even holds for OMFL on uniform metrics, and therefore w.r.t. the
Remark~\ref{rem:especial_case} also holds for the G-OMFL problem.
\end{remark}

More precisely, we prove the following main theorem.

\begin{theorem}\label{thm:UB_GUOMFL}
  Let \opt\ denote an optimal offline algorithm.
  There is a deterministic online algorithm \dga\ such that for all times $t \geq 0$,
  the total cost of \dga\ can be upper bounded as follows.
  \begin{itemize}
  \item If $\alpha=1$ and $\beta = \Omega\left(k+\frac{k}{\eps}\right)$
    for any $\eps>0$, it holds that
    \[
    \cost_t^\dga \leq (1+\eps)\cdot\cost_t^\opt + 
    O(k  \log k + \beta).
    \]
  \item If $\alpha=1$ and $\beta = \Omega\left(\frac{k \cdot \log
        k}{\log\log k}\right)$, then for every $\eps \geq \frac{\log \log
    k}{\log^{1-\delta}k}$ and any constant $0 < \delta \leq 1$, we obtain
    \[
    \cost_t^\dga \leq (1+\eps)\cdot\cost_t^\opt + 
    O\left(\beta\right).
    \]
  \end{itemize}
\end{theorem}

\para{Choosing \boldmath$\alpha>1$}
The results of the above
theorem all hold for $\alpha=1$, i.e., our algorithm is always forced
to move to a configuration that is optimal up to the additive term
$\beta$. Even if $\alpha$ is chosen to be larger than $1$, as long as
we want to guarantee a reasonably small multiplicative competitive
ratio (of order $o(k)$), an additive term of order $\Omega(k)$ is
unavoidable. In fact, in order to reduce the additive term to $O(k)$,
$\alpha$ has to be chosen to be of order $k^\delta$ for some constant
$\delta>0$. Note that in this case, the multiplicative competitive
ratio grows to at least $\alpha\gg1$. However, it might still be
desirable to choose $\alpha>1$. In that case, it can be shown that the
movement cost $\move_t^\dga$
only grows logarithmically with the optimal service cost
$\Service_t^*$ (where the basis of the logarithm is $\alpha$).
As an application, this, for example, allows being $(1+\eps)$-competitive for any constant $\eps>0$ against an objective function of the form
$\gamma\cdot \Service_t^\dga + \move_t^\dga$ even if $\gamma$ is
chosen of order $k^{-O(1)}$.

\subsubsection{Related Work}
\label{sec:intro_g-omfl_related}
The OMFL variant is a specific instance of the G-OMFL problem on uniform metrics, where the service cost function is the most basic. Recall that in Section~\ref{sec:intro_omfl_problem}, we defined the assignment cost for the OMFL problem as the distance between each request and its nearest facility. In a uniform metric, this means that the assignment cost is equal to the number of requests that are not co-located with any facility.

We have introduced a general service cost model for G-OMFL in Section~\ref{sec:intro_g-omfl_problem}. This model is similar to the one used by Hajiaghayi et al. \cite{hajiaghayi2003facility} for facility location, where the opening cost of a facility depends on the number of requests it serves. Another related generalization was made for the $k$-server problem by \cite{cote2008randomized,bansal2011polylogarithmic}, who defined a cost function for each subtree of an HST. They used an online algorithm for this variant, called the ``allocation problem'' that is defined on the uniform metrics, as a building block to design an online algorithm for $k$-server on the HST. The guarantees obtained by \cite{cote2008randomized} for the allocation problem on a two-point metric space allowed them to obtain a polylogarithmic-competitive algorithm for the $k$-server problem on a binary HST with sufficient separation. However, this algorithm is limited by the binary and well-separated structure of the HST. \cite{bansal2011polylogarithmic} extended \cite{cote2008randomized} to general HSTs and presented the first polylogarithmic-competitive algorithm for the $k$-server problem on arbitrary finite metrics with a competitive ratio of $O(\log^3 n \log^2 k \log \log n)$. This result is remarkable considering that the best deterministic upper bound for the $k$-server problem is $2k-1$ \cite{koutsoupias1995k} while there are no lower bounds better than $\Omega(\log k)$ \cite{fiat1991competitive} for the $k$-server problem. The randomized $k$-server conjecture states that the expected competitive ratio for the $k$-server is $O(\log k)$. 

\subsection{M-OMFL}
\label{sec:intro_m-omfl}
A naive algorithm for OMFL may not relocate any facilities and output the initial facility location as its solution. This may result in a high assignment cost, but no movement cost. However, this may not be desirable for companies that have to keep the total assignment cost below some threshold. There are several reasons why companies may have such a constraint on the assignment cost. One reason is budget limitation. Companies may have a fixed amount of money to spend on assigning requests to mobile facilities. Another reason is market competition. If the assignment cost is too high, it may affect the end users, making the company's services less attractive than those of similar companies that assign requests from closer facilities. For example, if a company serves end users from mobile facilities that are far away from them, it may incur a higher assignment cost, which could reduce their attractiveness. To address this issue, we define a variant of OMFL where any solution has to keep the total assignment cost below some threshold as a function of optimal assignment cost and at the same time the goal is to minimize the movement cost.

\subsubsection{Problem Defintion}
\label{sec:intro_m-omfl_problem}
The problem statement and the model definition for M-OMFL are the same as for OMFL, as stated in Section~\ref{sec:intro_omfl_problem}. The only difference is that the goal is to minimize the movement cost only, subject to a constraint that the total assignment cost does not exceed a given threshold for any algorithm that solves the problem, including an offline algorithm. The movement cost and the assignment cost are defined as in Section~\ref{sec:intro_omfl_problem}.
The threshold is defined as a function of the optimal
assignment cost. The problem is specified by two parameters $\alpha$ and
$\beta$ such that
\begin{equation}\label{eq:alphabetacondition_MMUOMFL}
\fpar \geq 1 \quad\text{and}\quad
\max\{\fpar-1,\spar\} \geq 1.
\end{equation}

\para{Feasible Configuration} We define a configuration $F$
to be feasible at time $t$ iff
\begin{equation}\label{eq:bounds_MMUOMFL}
	\Service_t(F) < \fpar \cdot \Service_t^* + \spar.
\end{equation}

\para{Feasible Solution} For a given algorithm \givenalg, we denote
the solution at time $t$ by $\algsol^\givenalg_t:=\{F^\givenalg(i): i \in
  [0,t]\}$ that is a sequence of facility configurations of algorithm \givenalg\, where $F^\givenalg(t)$ is the feasible configuration after reacting to
the arrival of request $t$ and where $F^\givenalg(0)=F_0$.
The assignment cost of an algorithm \givenalg\ at time $t$ is denoted by
$\Service_t^\givenalg := \Service_t(F^\givenalg(t))$.

The total cost of an algorithm, including an optimal offline algorithm, is the sum of the movement costs incurred by the algorithm. However, an optimal offline algorithm for M-OMFL cannot defer all the server movements until the end, as it can for OMFL and G-OMFL. It has to adjust the facility configuration whenever the condition in~\eqref{eq:bounds_MMUOMFL} is violated.

\subsubsection{Our Results}
\label{sec:intro_m-omfl_result}

We show that any deterministic online algorithm that solves an instance of
the M-OMFL problem in uniform metrics
necessarily has a competitive ratio of at least $\Omega(n)$.

\begin{theorem}\label{thm:LB_MMUOMFL}
  Assume that we are given parameters $\alpha$ and $\beta$ which
  satisfy \eqref{eq:alphabetacondition_MMUOMFL}. Then, for any online
  algorithm \on\ and for every $1\leq k<n$, there exists an execution and a time $t > 0$ such
  that the competitive ratio between the number of movements by \on\
  and the number of movements of an optimal offline algorithm \opt\
  is at least $n/2$. More precisely for all $\move_t^\opt>0$
  there is an execution such that $\move_t^\on \geq
  \frac{n}{2} \cdot \move_t^\opt$.
\end{theorem}

\subsubsection{Related Work}
\label{sec:intro_m-omfl_related}
Theorems~\ref{thm:LB_OMFL} and ~\ref{thm:UB_GUOMFL} demonstrate that if we aim to maintain the facilities in a configuration with an optimal assignment cost up to an additive $\beta$ for small $\beta$, we must pay a multiplicative competitive ratio of $\Omega(k)$. This is similar to the deterministic competitive ratio of $\Theta(k)$ for the $k$-server problem \cite{sleator1985amortized,koutsoupias1995k}. The M-OMFL is analogous to the $k$-server problem, where only the movement cost is minimized. When the distances between every pair of points are uniform, this variant becomes more similar to the paging problem. However, while the paging problem has a deterministic competitive ratio of $k$ \cite{manasse1990competitive}, we prove that the deterministic M-OMFL on uniform metrics (and hence on general metrics) has a lower bound of $n/2$, where $n$ is the number of metric points. Our lower bound analysis for M-OMFL is similar to the classic deterministic lower bound analysis for paging, particularly in how we compare the movements by the optimal offline algorithm and any online algorithm in each phase or interval. The randomized competitive ratio for the paging problem is $\Theta(\log k)$ \cite{fiat1991competitive}.

\section{G-OMFL: An Upper Bound Analysis}
\label{sec:G-OMFL}
In this section, we provide a proof of Theorem~\ref{thm:UB_GUOMFL}. The proof of Theorem~\ref{thm:LB_OMFL} is postponed to Section~\ref{sec:OMFL}, as it also holds for G-OMFL, considering Remark~\ref{rem:lowerbound}.
In Section~\ref{sec:algorithm_GUOMFL}, we introduce our online algorithm, the \emph{dynamic greedy allocation} (\DGA) denoted by \dga, for solving the G-OMFL problem. An overview of our analysis of \DGA\ is provided in Section~\ref{sec:intuition_GUOMFL}. The complete analysis of \DGA\ is presented in Section~\ref{sec:upperboundproof_GUOMFL}. In the following, whenever clear from the context, we omit the
superscript \dga\ in the algorithm-dependent quantities defined above.

\subsection{Algorithm Description}
\label{sec:algorithm_GUOMFL}
The goal of our algorithm is two-fold. On the one hand, we want to
guarantee that the service cost of \DGA\ is always within some
fixed bounds of the optimal service cost. On the other hand, we want
to achieve this while keeping the overall movement cost
low. Specifically, for two parameters $\alpha$ and $\beta$, where
\begin{equation}\label{eq:alphabetacondition_GUOMFL}
\fpar \geq 1 \quad\text{and}\quad
\max\{\fpar-1,\spar\} \geq 1.
\end{equation}
we guarantee that at all times
\begin{equation}\label{eq:algguarantee_GUOMFL}
  \Service_t < \alpha\cdot \Service_t^* + \beta,
\end{equation}
where $\Service_t$ denotes the total service cost of \DGA\ at
time $t$. Condition \eqref{eq:algguarantee_GUOMFL} is maintained in the most
straightforward greedy manner. Whenever after a new request arrives,
Condition \eqref{eq:algguarantee_GUOMFL} is not satisfied, \DGA\ greedily move
facilities until Condition \eqref{eq:algguarantee_GUOMFL} holds again. Hence, as long as
Condition \eqref{eq:algguarantee_GUOMFL} does not hold, \DGA\ moves a facility
that reduces the total service cost as much as possible. Our algorithm
stops moving any facilities as soon as the validity of
Condition \eqref{eq:algguarantee_GUOMFL} is restored.

Whenever \DGA\ moves a facility, it does a best possible move,
i.e., a move that achieves the best possible service cost
improvement. Thus, \DGA\ always moves a facility from a node
where removing a facility is as cheap as possible to a node where
adding a facility reduces the cost as much as possible. Therefore, for
each movement $m$, we have
\begin{eqnarray}\label{eq:movementsource_GUOMFL}
  \src_m & \in &
  \arg\min_{v\in V}
  \{\service_{v,\mtime_m}(\nfac_{v,m-1}-1)-\service_{v,\mtime_m}(\nfac_{v,m-1})\}\text{
    and}\\
  \label{eq:movementdest_GUOMFL}
  \dest_m & \in &
  \arg\max_{v\in V} \{\service_{v,\mtime_m}(\nfac_{v,m-1})-\service_{v,\mtime_m}(\nfac_{v,m-1}+1)\}.
\end{eqnarray}
Let $\mtime_m$ be the time of the $m$-$\mathit{th}$ movement and $\nfac_{v,m-1}$ be the number of facilities at node $v$ after the $(m-1)$-$\mathit{th}$ movement. As defined in Section~\ref{sec:intro_g-omfl_problem}, $\service_{v,t}(x)$ is the service cost at node $v$ of having $x$ facilities at time $t$. \eqref{eq:movementsource_GUOMFL} identifies a set of points where removing a facility causes the least increase in service cost. \eqref{eq:movementdest_GUOMFL} identifies a set of points where adding a facility causes the most decrease in service cost. DGA then moves a facility from a point in the first set to a point in the second set.

\subsection{Analysis Overview}
\label{sec:intuition_GUOMFL}
While the algorithm \DGA\ itself is quite simple, its analysis turns out
relatively technical. We thus first describe the key steps of the
analysis by discussing a simple case. 
We assume the classic cost model where on uniform metrics the service cost
at any node is equal to $0$ if there is at least one facility at the
node and the service cost is equal to the number of requests at the node,
otherwise. Further, we assume that we run \DGA\ with parameters $\alpha=1$ and $\beta=0$, i.e.
after each request arrives, \DGA\ moves to a configuration with optimal
service cost. Note that these parameter settings violate Condition
\eqref{eq:algguarantee_GUOMFL} and we therefore get a weaker bound
than the one promised by Theorem~\ref{thm:UB_GUOMFL}.

First, note that in the described simple scenario, \DGA\
clearly never puts more than one facility to the same node. Further,
whenever \DGA\ moves a facility from a node $u$ to a node $v$,
the overall service cost has to strictly decrease and thus, the number
of requests at node $v$ is larger than the number of requests at node
$u$. Consider some point in time $t$ and let
\[
	\ndem_{\min}(t):=\min\limits_{v\in V:\nfac_{v,t}=1}\ndem_{v,t}
\]
be the
minimum number of requests among the nodes $v$ with a facility at time
$t$. Hence, whenever at a time $t$, \DGA\ moves a facility
from a node $u$ to a node $v$, node $u$ has at least $\ndem_{\min}(t)$
requests and consequently, node $v$ has at least $\ndem_{\min}(t)+1$
requests. Further, if at some later time $t'>t$, the facility at node
$v$ is moved to some other node $w$, because \DGA\ always
removes a facility from a node with as few requests as possible, we
have $\ndem_{\min}(t')\geq\ndem_{\min}(t)+1$. Consequently, if in some
time interval $[t_1,t_2]$, there is some facility that is moved more
than once, we know that $\ndem_{\min}(t_1)<\ndem_{\min}(t_2)$. We partition time into phases, starting from time $0$. Each phase is a maximal time interval where no facility is moved more than once. (cf.\
Definition~\ref{def:phases_GUOMFL} in the formal analysis of \DGA).

The above argument implies that after each phase $\ndem_{\min}$
increases by at least one and therefore at any time $t$ in Phase $p$,
we have $\ndem_{\min}(t)\geq p-1$ and at the end of Phase $p$, we have
$\ndem_{\min}(t)\geq p$. In Section~\ref{sec:upperboundproof_GUOMFL}, the more
general form of this statement appears in
Lemma~\ref{lemma:phaseprogress_GUOMFL}. There, $\gamma_p$ is defined to be the
smallest service cost improvement of any movement in Phase $p$
($\gamma_p=1$ in the simple case considered here), and
Lemma~\ref{lemma:phaseprogress_GUOMFL} shows that $\ndem_{\min}$ grows by at least
$\gamma_p$ in Phase $p$.
Assume that at some time $t$ in Phase $p$, a facility is moved from a
node $u$ to a node $v$. Because node $u$ already had its facility at
the end of Phase $p-1$, we have $\ndem_{u,t}=\ndem_{\min}(t)\geq
p-1$. Consequently, at the end of Phase $p$, there is at least one
node (the source of the last movement) that has no facility and at
least $p-1$ requests. The corresponding (more technical) statement in
our general analysis appears in Lemma~\ref{lemma:manyrequests_GUOMFL}. 

We bound the total cost of \DGA\ and an optimal
offline algorithm from above and below, respectively, as a function of
the optimal service cost. Hence, the ratio between these two total
costs provides the desired competitive factor.  Our algorithm
guarantees that at all times, the service cost is within fixed bounds
of the optimal service cost (in the simple case here, the service cost
is always equal to the optimal service cost). Knowing that there are
nodes with many requests and no facilities, therefore allows to lower
bound the optimal service cost. In the general case, this is done by
 Lemma~\ref{lemma:betagrowth_GUOMFL} and Lemma~\ref{lemma:alphagrowth_GUOMFL}.
 In the simple case, considered here, as at the end of Phase $p$, there
are $k$ nodes with at least $p$ requests (the nodes that have facilities)
and there is at least one additional node with at least $p-1$
requests, we know that at the end of Phase $p$, the optimal service
cost is at least $p-1$. Consequently, \DGA\ (in the
simple case) pays exactly the optimal service cost (as mentioned
before, in the general case, the service cost is within fixed bounds
of the optimal service cost) and at most $(p-1)k$ as movement cost.
Hence, the total cost paid by \DGA\ is at most a factor
$k+1$ times the optimal service cost since the optimal service cost is
at least $p-1$. By choosing $\alpha$ slightly larger than
$1$ and a larger $\beta$ ($\beta\geq k$), \DGA\ becomes
lazier and one can show that the difference between the number of
movements of \DGA\ and the optimal service cost becomes significantly
smaller. Also note that by construction, the service cost of \DGA\ is
always at most $\alpha \Service_t^* + \beta$.

When analyzing \DGA, we mostly ignore the movement cost of an optimal offline algorithm. We only exploit the
fact that by the time \DGA\ decides to move a facility for the first
time, any other algorithm must also move at least one facility and
therefore the optimal offline cost becomes at least $1$. 

\subsection{Upper Bound Analysis}
\label{sec:upperboundproof_GUOMFL}
In the following, we show that how to upper bound the total cost
of our online algorithm denoted by \dga\ by a
function of the total cost of an optimal offline algorithm
\opt. Clearly, \DGA\ at all times $t\geq 0$ guarantees that
the service cost can be bounded as
\begin{equation}\label{eq:servicecompetitive_GUOMFL}
    \Service_t^\dga < \alpha \cdot \Service_t^* + \beta \leq \alpha \cdot \cost_t^\opt + \beta.
\end{equation}
In order to upper bound the total cost, it therefore suffices to
study how the movement cost $\move_t^\dga$ of \DGA\
grows as a function of the optimal offline algorithm cost. Let \opt\ be an
optimal offline algorithm and let $F^\opt(t)$ be the configuration of
\opt\ at time $t$. Recall that $\ch(F_0,F^\opt(t))$ denotes the total
number of movements required to move from the initial configuration to
configuration $F^\opt(t)$. We therefore have $\cost_t^\opt =
\Service_t^\opt + \move_t^\opt \geq \Service_t^* +
\ch(F_0,F^\opt(t))$. In order to upper bound $\move_t^\dga$ as a
function of $\cost_t^\opt$, we upper bound it as a function of
$\Service_t^* + \ch(F_0,F^\opt(t))$.

Instead of directly dealing with $\ch(F_0,F^\opt(t))$, we make
use of the fact that our analysis works for a general cost function
$\service$ satisfying the conditions given in
\eqref{eq:monotonicfac_GUOMFL}, \eqref{eq:monotonicdem_GUOMFL},
\eqref{eq:monotonicdifffac_GUOMFL}, and \eqref{eq:monotonicdiffdem_GUOMFL}. Given
an service cost function $\service$, consider a function $\service'$
which is defined as follows:
\[
\forall v\in V, \forall x\in \{0,\dots,k\}, \forall y\in
\mathbb{N}_0:
\service_v'(x,y) := \service_v(x,y) + \max\{0,f_{v}(0)-x\}
\]
where $f_{v}(t)$ is the number of facilities at time $t$ on node $v$. Clearly, $\service'$ also satisfies the conditions given in
\eqref{eq:monotonicfac_GUOMFL}, \eqref{eq:monotonicdem_GUOMFL},
\eqref{eq:monotonicdifffac_GUOMFL}, and \eqref{eq:monotonicdiffdem_GUOMFL}. In
addition, for any time $t$ and any configuration $F=\{(v,f_v) :
  v\in V\}$, we have
\begin{eqnarray}\nonumber
  \Service_t'(F)  & = & \sum_{v\in V} \service_v'(f_v,\ndem_{v,t})\\
  & = &\nonumber
\sum_{v\in V} \left(\service_v(f_v,\ndem_{v,t}) + 
  \max\{0,f_{v}(0)-f_v\}
\right)\\
& \stackrel{\eqref{eq:serverdistance_GUOMFL}}{=} & \label{eq:transformservocecost_GUOMFL}
\Service_t(F) + \ch(F_0, F)
\end{eqnarray}
where
$\Service_t'(F)$ refers to the total service cost w.r.t.\ the new cost 
function $\service'$. Hence, $\Service'_t(F)$ exactly measures the
sum of service cost and movement cost of a configuration $F$. Of course
now, in all our results, $\Service_t^*$ corresponds to the combination
of service and movement cost of an optimal configuration $F^*$. 

We are now going to analyze \DGA. In our analysis, we bound the total costs of
optimal offline algorithm \opt\ and online algorithm \dga\ from below and above,
respectively, as functions of optimal service cost and thus provide
the upper bound (competitive factor) promised in Theorem~\ref{thm:UB_GUOMFL}.
Hence we first go through calculating the optimal service cost.  

For the analysis of DGA, we
partition the movements into phases $p=1,2,\dots$, where
roughly speaking, a phase is a maximal consecutive sequence
of movements in which no facility is moved twice.
We use $\movement_p$ to denote the first movement of Phase $p$
(for $p\in\mathbb{N}$). In addition, we define $\srcalg_m$ and $\destalg_m$
to be the nodes involved in the $m$-th facility move, where we
assume that \dga\ moves a facility from node $\src_m$ to $\dest_m$.
Formally, the phases are defined as follows.
\begin{definition}[\bf{Phases}]\label{def:phases_GUOMFL}
  The movements are divided into phases $p=1,2,\dots$, where Phase $p$
  starts with movement $\movement_p$ and ends with movement
  $\movement_{p+1}-1$. We have $\movement_1=1$, i.e., the first phase
  starts with the first movement. Further for every $p>1$, we define
  \begin{equation}\label{eq:phasetime_GUOMFL}
    \movement_p := \min\{m>\movement_{p-1}\: :\:
      \exists\,m'\in[\movement_{p-1},m-1]\text{ s.t.\ } 
    \src_m=\dest_{m'}\}.
  \end{equation}
\end{definition}
For a Phase $p\geq1$, let $\lambda_p:=m_{p+1}-m_p$ be the number of
movements of Phase $p$. 

\subsubsection{Optimal Service Cost Analysis}
\label{sec:optservicecostlowerbound_GUOMFL}

The algorithm \DGA\ moves facilities in order to improve the service
cost. Throughout the rest of our analysis, we use $\mtimealg_m$ to
denote the time of the $m$-$\mathit{th}$ movement.
For a given movement $m$, we use $\gamma(m)>0$ to denote service
cost \emph{improvement} of $m$. Further, we use $F_0$ to denote the initial configuration of the $k$
facilities and for a given (deterministic) algorithm \givenalg, for
any $m\geq 1$, we let $F_m^\givenalg=\{(v,\nfac_{v,m}^\givenalg):v\in V\}$ be the
configuration of the $k$ facilities for \givenalg\ after $m$ facility movements (i.e., after $m$
server movements of \givenalg, node $v$ has $\nfac_{v,m}^\givenalg$
facilities). 

\begin{eqnarray}\nonumber
\gamma(m) & := & \ S_{\mtime_m}(F_{m-1}) - S_{\mtime_m}(F_m)\\
& \;= & \nonumber
\big(\service_{\dest_m,\mtime_m}(\nfac_{\dest_m,m-1})-\service_{\dest_m,\mtime_m}(\nfac_{\dest_m,m})\big)\\
&& \label{eq:improvement_GUOMFL}  \quad-\,
\big(\service_{\src_m,\mtime_m}(\nfac_{\src_m,m})-\service_{\src_m,\mtime_m}(\nfac_{\src_m,m-1})\big).
\end{eqnarray}

For each Phase $p$, we define the improvement $\gamma_p$ of $p$ and
the \emph{cumulative improvement} $\Gamma_p$ by Phase $p$ as follows
\begin{equation}\label{eq:phaseimprovement_GUOMFL}
  \gamma_p := \min_{m\in[m_p,m_{p+1}-1]} \gamma(m)\quad\text{and}\quad
  \Gamma_p := \sum_{i=1}^p \gamma_i,\quad\Gamma_0:=0,\ \gamma_0 := 0.
\end{equation}

We are now ready to prove our first technical lemma, which lower
bounds the cost of removing facilities from nodes with facilities (for all $v\in V$
such that $\nfac_v \geq 1$) at any point in the execution. The result of following lemma
implies that removing any facility of an optimal configuration during some Phase
$p$ increases the optimal service cost at least $\Gamma_{p-1}$
(and $\Gamma_p$ at end of Phase $p$) since
the facilities of an optimal configuration are located at places with maximum number
of requests.

\begin{lemma}\label{lemma:phaseprogress_GUOMFL}
  Let $m$ be a movement and, $F=\{(v,f_v):v\in V\}$ be the configuration
  of \DGA\ at any point in the execution after movement $m$
  and let $t\geq\mtime_m$ be the time at which the configuration $F$
  occurs. Then, for all times $t'\geq t$ and for all nodes $v\in V$,
  if $f_v>0$ it holds that
  \[
  	\service_{v,t'}(f_v-1) -
  	\service_{v,t'}(f_v)\geq \Gamma_{p-1},
  \]
  where $p$ is the phase in
  which movement $m$ occurs.
\end{lemma}
\begin{proof}
  We show that for each facility movement $m\in\mathbb{N}$ of \DGA, it holds that
  \begin{equation}\label{eq:aftermovement_GUOMFL}
    \forall v\in V\::\:
    \nfac_{v,m}>0\ \Longrightarrow\ 
    \service_{v,\mtime_m}(\nfac_{v,m}-1) -
    \service_{v,\mtime_m}(\nfac_{v,m})\geq \Gamma_{p-1},    
  \end{equation}
  where $p$ is the phase in which movement $m$ occurs (i.e., the claim
  of the lemma holds immediately after movement $m$). The lemma then
  follows because $(i)$ any configuration $\{(v,f_v):v\in V\}$ occurring
  after movement $m$ is the configuration $F_{m'}$ for some movement
  $m'\geq m$, $(ii)$ the values $\Gamma_{p-1}$ are monotonically
  increasing with $p$, and $(iii)$ by \eqref{eq:monotonicdiffdem_GUOMFL}, for all
  $v\in V$, the value $\service_{v,t}(f-1)-\service_{v,t}(f)$ is
  monotonically non-decreasing with $t$.

  It therefore remains to prove \eqref{eq:aftermovement_GUOMFL} for every
  $m$, where $p$ is the phase of movement $m$. We prove a slightly
  stronger statement. Generally, for a movement $m'$ and a Phase $p'$,
  let $\Dest_{p',m'}$ be the set of nodes that have received a new
  facility by some movement $m''\leq m'$ of Phase $p'$. Hence,
  \[
  \Dest_{p',m'}:=\{v\in V: \exists \text{ movement } m''\leq m'
    \text{ of Phase } p' \text{ s.t. } \dest_{m''}=v\}.
   \] 
    We show that in
  addition to \eqref{eq:aftermovement_GUOMFL}, it also holds that
  \begin{equation}\label{eq:extaftermovement_GUOMFL}
    \forall v\in \Dest_{p,m}\::\:
    \nfac_{v,m}>0\ \Longrightarrow\ 
    \service_{v,\mtime_m}(\nfac_{v,m}-1) -
    \service_{v,\mtime_m}(\nfac_{v,m})\geq \Gamma_{p}.
  \end{equation}
  We prove \eqref{eq:aftermovement_GUOMFL} and \eqref{eq:extaftermovement_GUOMFL}
  together by using induction on $m$.

  \smallskip
  \noindent\textbf{Induction Base \boldmath($m=1$):} The first
  movement occurs in Phase $1$. By \eqref{eq:phaseimprovement_GUOMFL},
  $\Gamma_0=0$ and by \eqref{eq:monotonicfac_GUOMFL}, we also have
  $\service_{v,t}(f-1)-\service_{v,t}(f)\geq 0$ for all times $t\geq
  0$, all nodes $v\in V$, and all $f\geq 1$. Inequality
  \eqref{eq:aftermovement_GUOMFL} therefore clearly holds for $m=1$. It
  remains to show that also \eqref{eq:extaftermovement_GUOMFL} holds for
  $m=1$. We have $\Dest_{1,1}=\{\dest_1\}$ and showing
  \eqref{eq:extaftermovement_GUOMFL} for $m=1$ therefore reduces to showing
  that
  $\service_{\dest_1,\mtime_1}(\nfac_{\dest_1,1}-1)-\service_{\dest_1,\mtime_1}(\nfac_{\dest_1,1})\geq
  \Gamma_1=\gamma_1$, which follows directly from
  \eqref{eq:improvement_GUOMFL} and \eqref{eq:phaseimprovement_GUOMFL}.

  \smallskip
  \noindent\textbf{Induction Step \boldmath($m>1$):} We first show
  that Inequalities \eqref{eq:aftermovement_GUOMFL} and
  \eqref{eq:extaftermovement_GUOMFL} hold immediately before movement $m$ and
  thus,
  \begin{eqnarray}\label{eq:beforemovement_GUOMFL}
    \forall v\in V & : &
    \nfac_{v,m-1}>0\ \Rightarrow\ 
    \service_{v,\mtime_m}(\nfac_{v,m-1}-1) -
    \service_{v,\mtime_m}(\nfac_{v,m-1})\geq \Gamma_{p-1},\\
    \label{eq:extbeforemovement_GUOMFL}
    \forall v\in \Dest_{p,m-1} & : &
    \nfac_{v,m-1}>0\ \Rightarrow\ 
    \service_{v,\mtime_m}(\nfac_{v,m-1}-1) -
    \service_{v,\mtime_m}(\nfac_{v,m-1})\geq \Gamma_{p}.
  \end{eqnarray}
  If $m$ is not the first movement of Phase $p$, Inequalities
  \eqref{eq:beforemovement_GUOMFL} and \eqref{eq:extbeforemovement_GUOMFL} follow
  directly from the induction hypothesis (for $m-1$) and from
  \eqref{eq:monotonicdiffdem_GUOMFL}. Let us therefore assume that $m$ is the
  first movement of Phase $p$. Note that in this case
  $\Dest_{p,m-1}=\emptyset$ and \eqref{eq:extbeforemovement_GUOMFL} therefore
  trivially holds. Because $m>1$, we know that in this case $p\geq
  2$. From the induction hypothesis and from
  \eqref{eq:monotonicdiffdem_GUOMFL}, we can therefore conclude that for
  every node $v\in \Dest_{p-1,m-1}$ (every node $v$ that is the destination of some
  facility movement in Phase $p-1$), we have
  $\service_{v,\mtime_m}(\nfac_{v,m-1}-1) -
  \service_{v,\mtime_m}(\nfac_{v,m-1})\geq \Gamma_{p-1}$. Note that
  for all these nodes, we have $\nfac_{v,m-1}>0$. Because $m$ is the
  first movement of Phase $p$, Definition~\ref{def:phases_GUOMFL} implies
  that $\src_m\in\Dest_{p-1,m-1}$. Applying \eqref{eq:movementsource_GUOMFL},
  we get that for all $v\in V$,
  $\service_{v,\mtime_m}(\nfac_{v,m-1}-1) -
  \service_{v,\mtime_m}(\nfac_{v,m-1})\geq
  \service_{\src_m,\mtime_m}(\nfac_{\src_m,m-1}-1) -
  \service_{\src_m,\mtime_m}(\nfac_{\src_m,m-1})\geq \Gamma_{p-1}$ and
  therefore \eqref{eq:beforemovement_GUOMFL} also holds if $m\geq2$ is the
  first movement of some phase.

  We can now prove \eqref{eq:aftermovement_GUOMFL} and
  \eqref{eq:extaftermovement_GUOMFL}. For all nodes
  $v\notin\{\src_m,\dest_m\}$, we have $\nfac_{v,m}=\nfac_{v,m-1}$
  and we further have $\Dest_{p,m}=\Dest_{p,m-1}\cup
  \{\dest_m\}$. For $v\notin\{\src_m,\dest_m\}$,
  \eqref{eq:aftermovement_GUOMFL} and \eqref{eq:extaftermovement_GUOMFL} therefore
  directly follow from \eqref{eq:beforemovement_GUOMFL} and
  \eqref{eq:extbeforemovement_GUOMFL}, respectively. For the two nodes
  involved in movement $m$, first note that $\src_m\notin \Dest_{p,m-1}$. It therefore suffices
  to show that
  \begin{eqnarray}\label{eq:Gammainductionstepsrc_GUOMFL}
    \nfac_{\src_m,m}=0\quad \text{or}\quad \service_{\src_m,\mtime_m}(\nfac_{\src_m,m}-1) -
    \service_{\src_m,\mtime_m}(\nfac_{\src_m,m}) & \geq &
    \Gamma_{p-1},\\
    \label{eq:Gammainductionstepdest_GUOMFL}
    \text{as well as}\quad
    \service_{\dest_m,\mtime_m}(\nfac_{\dest_m,m}-1) -
    \service_{\dest_m,\mtime_m}(\nfac_{\dest_m,m}) & \geq &
    \Gamma_{p}.    
  \end{eqnarray}
  We have $\nfac_{\src_m,m} = \nfac_{\src_m,m-1}-1$ and
  $\nfac_{\dest_m,m}=\nfac_{\dest_m,m-1}+1$. Inequality
  \eqref{eq:Gammainductionstepsrc_GUOMFL} therefore directly follows from
  \eqref{eq:beforemovement_GUOMFL} and \eqref{eq:monotonicdifffac_GUOMFL}.
  For \eqref{eq:Gammainductionstepdest_GUOMFL}, we have
  \begin{eqnarray*}
    \lefteqn{\service_{\dest_m,\mtime_m}(\nfac_{\dest_m,m}-1) -
    \service_{\dest_m,\mtime_m}(\nfac_{\dest_m,m})}\qquad\\ 
    & \stackrel{\eqref{eq:improvement_GUOMFL}}{=} & 
    \service_{\src_m,\mtime_m}(\nfac_{\src_m,m}) -
    \service_{\src_m,\mtime_m}(\nfac_{\src_m,m}-1) + \gamma(m)\\
    &
    \stackrel{\eqref{eq:beforemovement_GUOMFL}}{\geq}
    &
    \Gamma_{p-1}+ \gamma(m) \ \
    \stackrel{\eqref{eq:phaseimprovement_GUOMFL}}{\geq}\ \ \Gamma_p. 
  \end{eqnarray*}
  This completes the proof of \eqref{eq:aftermovement_GUOMFL} and
  \eqref{eq:extaftermovement_GUOMFL} and thus the proof of the lemma. 
\end{proof}

For each phase number $p$, let $\ptime_p:=\mtime_{m_p}$ be the time of
the the first movement $m_p$ of Phase $p$.  Before continuing, we give
lower and upper bounds on $\gamma_p$, the improvement of Phase
$p$. For all $p\geq 1$, we define
\begin{equation}\label{eq:eta_GUOMFL}
    \eta_p := (\alpha-1)\cdot \Service_{\ptime_p}^* + \beta.
\end{equation}

\begin{lemma}\label{lemma:gammabound_GUOMFL}
  Let $m$ be a movement of Phase $p$ and let $F^* \in \arg\min\limits_{F}\Service_t(F)$
  be the optimal configuration at time
  $\mtime_m$. We then have
  \[
  \frac{\eta_p}{\ch(F_{m-1},F^*)} \leq \gamma(m) \leq \eta_{p+1}.
  \]
\end{lemma}
\begin{proof}
  For the upper bound, observe that we have
  \[
  \gamma(m)\leq \Service_{\mtime_m}(F_{m-1})-\Service_{\mtime_m}^*
  \]
  as clearly the service cost cannot be improved by a larger amount. Because at all
  times $t$, \DGA\ keeps the service cost below
  $\alpha\Service_t^*+\beta$, we have $\Service_{\mtime_m-1}(F_{m-1})
  < \alpha\Service_{\mtime_m-1}^*+\beta
  \leq\alpha\Service_{\mtime_m}^*+\beta$. The upper bound on $\gamma(m)$ follows from
  \eqref{eq:eta_GUOMFL} and because
  $\Service_{\mtime_m}^*\leq\Service_{\ptime_{p+1}}^*$.
    
    For the lower bound on $\gamma(m)$, we need to prove that
    $\ch(F_{m-1},F^*)\geq \eta_p/\gamma(m)$. Because \DGA\
    moves a facility at time $\mtime_m$, we know that
    $\Service_{\mtime_m}(F_{m-1})\geq\alpha \Service_{\mtime_m}(F^*) +
    \beta$ and applying the \eqref{eq:eta_GUOMFL} of $\eta_p$, we
    thus have $\Service_{\mtime_m}(F_{m-1})-\Service_{\mtime_m}(F^*)
    \geq \eta_p$. Intuitively, we have $\ch(F_{m-1},F^*)\geq
    \eta_p/\gamma(m)$ because \DGA\ always chooses the best
    possible movement and thus every possible movement improves the
    overall service cost by at most $\gamma(m)$. Thus, the number of
    movements needs to get from $F_{m-1}$ to an optimal configuration $F^*$
    has to be at least $\eta_p/\gamma(m)$. For a formal argument,
    assume that we are given a sequence of $\ell:=\ch(F_{m-1},F^*)$
    movements that transform configuration $F_{m-1}$ into configuration
    $F^*$. For $i\in[\ell]$, assume that the $i$-$\mathit{th}$ of
    these movements moves a facility from node $u_i$ to node
    $v_i$. Further, for any $i\in[\ell]$ let $f_i$ be the number of
    facilities at node $u_i$ and let $f_i'$ be the number of
    facilities at node $v_i$ before the $i$-$\mathit{th}$ of these
    movements. Because the sequence of movements is minimal to get
    from $F_{m-1}$ to $F^*$, we certainly have $f_i\leq f_{u_i,m-1}$
    and $f_i'\geq f_{v_i,m-1}$. For the service cost improvement
    $\gamma$ of the $i$-$\mathit{th}$ of these movements, we therefore
    obtain
  \begin{eqnarray*}
    \gamma & = & \big(\service_{v_i,\mtime_m}(f_i') -
    \service_{v_i,\mtime_m}(f_i'+1)\big) -
    \big(\service_{u_i,\mtime_m}(f_i-1)-\service_{u_i,\mtime_m}(f_i)\big)\\
    & \stackrel{\eqref{eq:monotonicdifffac_GUOMFL}}{\leq} &
    \big(\service_{v_i,\mtime_m}(f_{v_i,m-1}) -
    \service_{v_i,\mtime_m}(f_{v_i,m-1}+1)\big)\\ 
    && \quad -\, \big(\service_{u_i,\mtime_m}(f_{u_i,m-1}-1)-\service_{u_i,\mtime_m}(f_{u_i,m-1})\big)\\
    & \leq & \gamma(m).
  \end{eqnarray*}
  The last inequality follows from
  \eqref{eq:movementsource_GUOMFL},\eqref{eq:movementdest_GUOMFL}, and
  \eqref{eq:improvement_GUOMFL}. As the sum of the $\ell$ service cost
  improvements has to be at least $\eta_p$, we obtain
  $\ell=\ch(F_{m-1},F^*)\geq \eta_p/\gamma(m)$ as claimed.
\end{proof}

We can now lower bound the distribution of requests at the time of each
movement.
\begin{lemma}\label{lemma:manyrequests_GUOMFL}
  Let $m$ be a movement of Phase $p$ (for $p\geq 1$). Then, there are
  integers $\psi_v\geq 0$ for all nodes $v\in V$ such that 
  \begin{align*}
  &\sum_{v\in V} \psi_v \geq k + \frac{\eta_p}{\gamma(m)}
  \quad \text{and}\\
  \forall t\geq\mtime_m \:\forall v\in V : & \ \psi_v>0 \Longrightarrow
  \service_{v,t}(\psi_v-1) - \service_{v,t}(\psi_v) \geq \Gamma_{p-1}.
  \end{align*}
\end{lemma}
\begin{proof}
  It suffices to prove the statement for $t=\mtime_m$. For larger $t$,
  the claim then follows from \eqref{eq:monotonicdiffdem_GUOMFL}. Consider an
  optimal configuration
  \[
  F^*=\{(v,f_v^*):v\in V\}
  \]
  at the time $\mtime_m$ of movement
  $m$. Let us further consider the configuration $F_{m-1}$ of \DGA\
  immediately before movement $m$. Consider a pair of nodes $u$ and
  $v$ such that $f_u^*>f_{u,m-1}$ and $f_{v,m-1}>f_v^*$. By the
  optimality of $F^*$, we have
  \begin{equation}\label{eq:optvsalg_GUOMFL}
    \service_{u,\mtime_m}(f_u^*-1)-\service_{u,\mtime_m}(f_u^*) \geq
    \service_{v,\mtime_m}(f_{v,m-1}-1)-\service_{v,\mtime_m}(f_{v,m-1}).
  \end{equation}
  Otherwise, moving a facility from $u$ to $v$ would (strictly)
  improve the configuration $F^*$. By Lemma~\ref{lemma:phaseprogress_GUOMFL}, we
  have
  $\service_{v,\mtime_m}(f_{v,m-1}-1)-\service_{v,\mtime_m}(f_{v,m-1})\geq
  \Gamma_{p-1}$ for all nodes $v$ for which $f_{v,m-1}>0$. Together
  with \eqref{eq:optvsalg_GUOMFL}, for all $v\in V$ for which
  $\max\{f_{v,m-1},f_v^*\}>1$, we obtain
  \begin{equation}\label{eq:optalgcombined_GUOMFL}
    \service_{v,\mtime_m}(\max\{f_{v,m-1},f_v^*\}-1)-
    \service_{v,\mtime_m}(\max\{f_{v,m-1},f_v^*\})\geq
    \Gamma_{p-1}.
  \end{equation}
  To prove the lemma, it therefore suffices to show that $\sum_{v\in
    V} \max\{f_{v,m-1},f_v^*\} \geq k+\eta_p/\gamma(m)$, as we can
  then set $\psi_v:=\max\{f_{v,m-1},f_v^*\}$ and
  \eqref{eq:optalgcombined_GUOMFL} implies the claim of the lemma. By
  \eqref{eq:serverdistance_GUOMFL}, we have
  \[
  \sum_{v\in V}\max\{f_{v,m-1},f_v^*\} = k + \sum_{v\in
    V}\max\{0,f_v^*-f_{v,m-1}\} = k+\ch(F_{m-1},F^*).
  \]
  We therefore need that $\ch(F_{m-1},F^*)\geq \eta_p/\gamma(m)$,
  which follows from Lemma~\ref{lemma:gammabound_GUOMFL}.
\end{proof}

In the next lemma, we derive a lower bound on $\Service_{\ptime_p}^*$,
the service cost of optimal configuration when Phase $p$ starts.
For each Phase $p\geq1$, we first define $\ServiceLB{p}$ as
follows. 
\begin{equation}\label{eq:ServiceLB_GUOMFL}
  \text{For } p \geq 3: \ServiceLB{p} :=
  \left(1+ (\alpha-1)\frac{\gamma_{p-2}}{\gamma_{p-1}}\right)\cdot
  \ServiceLB{p-1}
  + \frac{\gamma_{p-2}}{\gamma_{p-1}}\beta,\text{ and }
  \ServiceLB{1}:=\ServiceLB{2} := 1.
\end{equation}

\begin{lemma}\label{lemma:ServiceLB_GUOMFL}
  For all $p\geq 1$, we have $\Service_{\ptime_p}^* \geq \ServiceLB{p}$.
\end{lemma}
\begin{proof}
  We prove the lemma by induction on $p$.

  \smallskip
  \noindent\textbf{Induction Base \boldmath($p=1,2$):}
  Using \eqref{eq:transformservocecost_GUOMFL}
  we have $\Service_{\ptime_1}^* \geq 1$
  and since $\ServiceLB{1}=\ServiceLB{2}=1$, we get
  $\Service_{\ptime_2}^*\geq\Service_{\ptime_1}^* \geq
  \ServiceLB{2}=\ServiceLB{1}$.

  \smallskip
  \noindent\textbf{Induction Step \boldmath($p>2$):} We use the
  induction hypothesis to assume that the claim of the lemma is true
  up to Phase $p$ and we prove that it also holds for Phase
  $p+1$. Therefore by the induction hypothesis, for all $i \in [p]$,
  \begin{equation}\label{eq:indhypserviceLB_GUOMFL}
  	\Service_{\ptime_p}^* \geq \ServiceLB{p}.
  \end{equation}
  For all $i\in[p]$, we define $\etaLB{i}:=
  (\alpha-1)\ServiceLB{i}+\beta$ and
  $\delta_i:=\max\big\{\frac{\etaLB{i+1}}{\gamma_{i+1}},\cdots,\frac{\etaLB{p}}{\gamma_{p}}\big\}$.
  As a consequence of \eqref{eq:eta_GUOMFL} and \eqref{eq:indhypserviceLB_GUOMFL},
  we get that $\eta_i\geq \etaLB{i}$ for all $i\in [p]$. In the
  following, let $p'\in[2,p]$ be some phase.
  Lemma~\ref{lemma:manyrequests_GUOMFL} implies that after the last movement $m$
  of Phase $p'$, there are non-negative integers $\psi_v$ (for $v\in
  V$) such that $\sum_{v\in V}\psi_v\geq k+\eta_{p'}/\gamma_{p'}\geq
  \etaLB{p'}/\gamma_{p'}$ and for all times $t\geq \mtime_m$, for all
  $v\in V$ for which $\psi_v>0$,
  $\service_{v,t}(\psi_v-1)-\service_{v,t}(\psi_v)\geq
  \Gamma_{p'-1}$. As there are only $k$ facilities for any feasible
  configuration $F=\{(v,f_v)\}$, we have $\sum_{v\in V} f_v=k$ and
  therefore $\sum_{v\in V}(\psi_v-f_v)\geq
  \etaLB{p'}/\gamma_{p'}$. For any $v\in V$ for which $\psi_v>f_v$, by
  using \eqref{eq:monotonicdifffac_GUOMFL}, we get $\service_{v,t}(f_v)\geq
  (\psi_v-f_v)\Gamma_{p'-1}$. Hence, after the last movement of Phase
  $p'$, for any feasible configuration $F$, we have $S_t(F)\geq S_t^* \geq
  \frac{\etaLB{p'}}{\gamma_{p'}}\Gamma_{p'-1}$. At the beginning of
  Phase $p+1$ (for $p\geq 2$), the total optimal service cost therefore is
  \begin{equation}\label{eq:serviceLBnextphase_GUOMFL}
    \Service_{\ptime_{p+1}}^* \geq \max_{p'\in[2,p]}
    \frac{\etaLB{p'}}{\gamma_{p'}} \Gamma_{p'-1} \geq
    \delta_{p-1}\Gamma_{p-1} +
    \sum_{i=1}^{p-2}(\delta_i-\delta_{i+1})\cdot\Gamma_i =
    \sum_{i=1}^{p-1}\gamma_i\cdot\delta_i.
  \end{equation}  

  We define $\zeta_i$ for all $i \in [3,p]$ as follows: 
  \begin{equation}\label{eq:zetaLB_GUOMFL}
    \zeta_i := \sum_{j=1}^{i-2} \gamma_j \cdot \delta_j.
  \end{equation}
  Using the definition of $\delta_i$, we thus have
  \[
  \zeta_{p+1} = \zeta_{p} + \gamma_{p-1} \delta_{p-1}=\zeta_{p} + \etaLB{p}\frac{\gamma_{p-1}}{\gamma_{p}}.
  \]
  Considering the definition of $\etaLB{i}$ we get
  \[
  \zeta_{p+1} = \zeta_{p} \cdot \left(1 + (\alpha-1)\frac{\gamma_{p-1}}{\gamma_{p}}\right) + \spar \cdot \frac{\gamma_{p-1}}{\gamma_{p}}.
  \]
  We therefore have $\zeta_{p+1}=\ServiceLB{p+1}$ directly from
  \eqref{eq:ServiceLB_GUOMFL} and thus the claim of the lemma follows.
\end{proof}

In order to explicitly lower bound the optimal service cost after $p$
phases, we need the following technical statement.

\begin{lemma}\label{lemma:fractionsum_GUOMFL}
  Let $\ell\geq 2$ be an integer and consider a sequence
  $c_1,c_2,\dots,c_\ell>0$ of $\ell$ positive real numbers and let
  $c_{\max}=\max\limits_{i\in[\ell]} c_i$ and $c_{\min}=\min\limits_{i\in[\ell]}
  c_i$. Further, let $\lambda\geq0$ be an arbitrary non-negative real
  number. We have
  \begin{align*}
    \mathrm{(I)}\quad &  \sum_{i=2}^\ell \frac{c_{i-1}}{c_i} \geq (\ell-1)\cdot
    \left(\frac{c_{\min}}{c_{\max}}\right)^{\frac{1}{\ell-1}},\\
    \mathrm{(II)}\quad  & \prod_{i=2}^{\ell} 
    \left(1+\lambda\frac{c_{i-1}}{c_i}\right) \geq
    \left(1+\lambda\left(\frac{c_{\min}}{c_{\max}}\right)^{\frac{1}{\ell-1}}\right)^{\ell-1}.
  \end{align*}
\end{lemma}
\begin{proof}
  The first part of the claim follows from the means inequality (the
  fact that the arithmetic mean is larger than or equal to the
  geometric mean). In the following, we nevertheless directly prove
  both parts together. We let
  $\vect{x}=(x_1,\dots,x_\ell)\in\mathbb{R}^\ell$ be a vector $\ell$
  real variables and we define multivariate functions
  $f(\vect{x}):\mathbb{R}^{\ell}\to\mathbb{R}$ and
  $g(\vect{x}):\mathbb{R}^{\ell}\to\mathbb{R}$ as follows:
  \[
  f(\vect{x}) :=
  \sum_{i=2}^{\ell}\frac{x_{i-1}}{x_i}\quad\text{and}\quad
  g(\vect{x}) :=
  \prod_{i=2}^{\ell}\left(1+\lambda\frac{x_{i-1}}{x_i}\right).
  \]
  We further define $X\subset\mathbb{R}^\ell$ as
  $X:=\{(z_1,\dots,z_\ell)\in\mathbb{R}^\ell\, \lvert \ \forall
    i\in[\ell]:c_{\min}\leq z_i\leq c_{\max}\}$. We need to show that
  for $\vect{x}\in X$, $f(\vect{x})$ and $g(\vect{x})$ are lower
  bounded by the right-hand sides of Inequalities (I) and (II) above,
  respectively. Note that $X$ is a closed subset of $\mathbb{R}^\ell$
  and because $c_{\min}>0$, both functions $f(\vect{x})$ and
  $g(\vect{x})$ are continuous when defined on $X$. The minimum for
  $\vect{x} \in X$ is therefore well-defined for both $f(\vect{x})$
  and $g(\vect{x})$. We show that both $f(\vect{x})$ and $g(\vect{x})$
  attain their minimum for 
  \[
  \vect{x^*} := (x_1^*,\dots,x_\ell^*),\quad\text{where }
  \forall i\in[\ell] : x_i^* = c_{\min}\cdot 
  \left(\frac{c_{\max}}{c_{\min}}\right)^{\frac{i-1}{\ell-1}}.
  \]
  Note that $\vect{x^*}$ is the unique configuration $\vect{x}\in X$ to the
  following system of equations
  \begin{equation}\label{eq:stationary_GUOMFL}
    x_1=c_{\min},\quad x_\ell=c_{\max},\quad
    \forall i\in\{2,\dots,\ell-1\}:
    x_i\in\frac{x_{i-1}}{x_i} = \frac{x_i}{x_{i+1}}.
  \end{equation}
  Because we know that $\min\limits_{\vect{x}\in X}f(\vect{x})=f(\vect{x^*})$
  and $\min\limits_{\vect{x}\in X} g(\vect{x})=g(\vect{x^*})$, it is
  therefore sufficient to show that for any $\vect{y}\in X$ that does
  not satisfy \eqref{eq:stationary_GUOMFL}, $f(\vect{y})$ and $g(\vect{y})$
  are not minimal. Let us therefore consider a vector
  $\vect{y}=(y_1,\dots,y_\ell)\in X$ that does not satisfy
  \eqref{eq:stationary_GUOMFL}. First note that both $f(\vect{x})$ and
  $g(\vect{x})$ are strictly monotonically increasing in $x_1$ and
  strictly monotonically decreasing in $x_\ell$. If either
  $y_1>c_{\min}$ or $y_\ell<c_{\max}$, it is therefore clear that
  $f(\vect{y})$ and $g(\vect{y})$ are both not minimal (over $X$). Let
  us therefore assume that $y_1=c_{\min}$ and $y_\ell=c_{\max}$. From
  the assumption that $\vect{y}$ does not satisfy
  \eqref{eq:stationary_GUOMFL}, we then have an $i_0\in\{2,\dots,\ell-1\}$
  for which $\frac{y_{i_0-1}}{y_{i_0}}\neq \frac{y_{i_0}}{y_{i_0+1}}$
  and thus $y_{i_0} \neq \sqrt{y_{i_0-1}y_{i_0+1}}$. We define a new
  vector $\vect{y'}=(y_1',\dots,y_\ell')\in X$ as follows. We have
  $y_{i_0}'=\sqrt{y_{i_0-1}y_{i_0+1}}$ and $y_i'=y_i$ for all $i\neq
  i_0$ and we show that $f(\vect{y'})<f(\vect{y})$ and
  $g(\vect{y'})<g(\vect{y})$. Define
  \[
  C:=\prod_{i\in[2,\ell]\setminus\{i_0,i_0+1\}}\left(1+\lambda\frac{y_{i-1}}{y_i}\right).
  \] 
  We then have
  \begin{align*}
    f(\vect{y})-f(\vect{y'}) & =
    \left(\frac{y_{i_0-1}}{y_{i_0}} +
      \frac{y_{i_0}}{y_{i_0+1}}\right)-
    \left(\frac{y_{i_0-1}}{y_{i_0}'} + \frac{y_{i_0}'}{y_{i_0+1}}\right)\\
    g(\vect{y})-g(\vect{y'}) & =
    \left[
      \left(1+\lambda\frac{y_{i_0-1}}{y_{i_0}}\right)\cdot      
      \left(1+\lambda\frac{y_{i_0}}{y_{i_0+1}}\right)-
      \left(1+\lambda\frac{y_{i_0-1}}{y_{i_0}'}\right)\cdot      
      \left(1+\lambda\frac{y_{i_0}'}{y_{i_0+1}}\right)
    \right]\cdot C\\
    & =
    \left[
      \left(\frac{y_{i_0-1}}{y_{i_0}} +
        \frac{y_{i_0}}{y_{i_0+1}}\right)-
      \left(\frac{y_{i_0-1}}{y_{i_0}'} + \frac{y_{i_0}'}{y_{i_0+1}}\right)
    \right]\cdot\lambda C.
  \end{align*}
  Note that $\lambda\geq 0$ and $C>0$. In both cases, we therefore
  need to show that
  \begin{equation}\label{eq:_GUOMFL}
    \forall y_{i_0}\in[c_{\min},c_{\max}]\setminus
    \{\sqrt{y_{i_0-1}y_{i_0+1}}\} :
    \left(\frac{y_{i_0-1}}{y_{i_0}} +
      \frac{y_{i_0}}{y_{i_0+1}}\right) >
    \left(\frac{y_{i_0-1}}{y_{i_0}'} + \frac{y_{i_0}'}{y_{i_0+1}}\right).
  \end{equation}
  This follows because the function
  $h:[c_{\min},c_{\max}]\to\mathbb{R}$,
  $h(z):=\frac{y_{i_0-1}}{z}+\frac{z}{y_{i_0+1}}$ is strictly convex
    for $z\in[c_{\min},c_{\max}]$ and it has a stationary point at
    $z=\sqrt{y_{i_0-1}y_{i_0+1}}\in[c_{\min},c_{\max}]$.
\end{proof}

As long as $(\alpha-1)\Service_{\ptime_p}^*<\beta$, the effect of the
$(\alpha-1)\Service_{\ptime_p}^*$-term on $\eta_p$ (and thus of the
$\alpha S_t^*$ term in \eqref{eq:algguarantee_GUOMFL} is relatively
small. Let us therefore first analyze how the service cost grows by
just considering terms that depends on $\beta$ (and not on
$\alpha$). 

\begin{lemma}\label{lemma:betagrowth_GUOMFL}
  For all $p\geq 3$, we have
  \[
  \Service_{\ptime_p}^* \geq \min\{
    \frac{\beta}{\alpha-1},\ 
    \beta\cdot(p-2)\cdot
  (2k)^{-\frac{1}{p-2}}\}.
  \]
\end{lemma}
\begin{proof}
  Assume that $\Service_{\ptime_p}^*<\beta/(\alpha-1)$ as otherwise
  the claim of the lemma is trivially true.  By
  Lemma~\ref{lemma:ServiceLB_GUOMFL}, using $\alpha\geq 1$, for all $p\geq 3$, we
  get $\ServiceLB{p} \geq  \ServiceLB{p-1} +
  \frac{\gamma_{p-2}}{\gamma_{p-1}}\beta$. Plugging in
  $ \ServiceLB{2} \geq 0$, induction on $p$ therefore gives
  \begin{equation}\label{eq:additivesum_GUOMFL}
      S_{\ptime_p}^* \geq  \ServiceLB{p} \geq
      \beta\cdot\sum_{i=2}^{p-1}\frac{\gamma_{i-1}}{\gamma_i}
  \end{equation}
  for all $p\geq 3$. We define
  $\gamma_{\min}=\min\{\gamma_1,\dots,\gamma_{p-1}\}$ and
  $\gamma_{\max}=\max\{\gamma_1,\dots,\gamma_{p-1}\}$. By
  Lemma~\ref{lemma:gammabound_GUOMFL} and because $\eta_1\leq\dots\leq\eta_{p-1}$,
  we have $\gamma_{\min}\geq \eta_1/k$ and $\gamma_{\max}\leq
  \eta_{p}$. From $\alpha\geq 1$ and \eqref{eq:eta_GUOMFL}, we
  have $\eta_1\geq (\alpha-1)+\beta$ since we know $\Service_{\ptime_p}^* \geq 1$ for $p\geq 1$
  regarding to \eqref{eq:transformservocecost_GUOMFL}. Further, we have
  $\eta_{p}=(\alpha-1)\Service_{\ptime_p}^*+\beta<2\beta$. We
  therefore have $\gamma_{\min}\geq[(\alpha-1)+\beta]/k$
  and $\gamma_{\max} < 2\beta$ and thus
  \[
  \frac{\gamma_{\min}}{\gamma_{\max}}\ \geq\
  \frac{(\alpha-1)+\beta}{2k\beta}
  \ \stackrel{\eqref{eq:alphabetacondition_GUOMFL}}{\geq}\
  \frac{\max\{\beta,1\}}{2k\beta}
  \ \geq\ \frac{1}{2k}.
  \]
  The lemma now follows from \eqref{eq:additivesum_GUOMFL} and from
  Inequality (I) of Lemma~\ref{lemma:fractionsum_GUOMFL}.
\end{proof}

On the other hand, as soon as $\Service_{\ptime_p}^* >
\max\{1,\frac{\beta}{\alpha-1}\}$, the effect of the
$\beta$-term in \eqref{eq:algguarantee_GUOMFL} becomes relatively small. As a
second case, therefore, we analyze how the service cost grows by just
considering terms that depends on $\alpha$ (and not on $\beta$).

\begin{lemma}\label{lemma:alphagrowth_GUOMFL}
  Let $p_0\geq 2$ be a phase for which $\ServiceLB{p_0}\geq
  \ServiceLB{p_0-1}\geq
  S_0:=\max\big\{1,\frac{\beta}{\alpha-1}\big\}$. For any phase
  $p>p_0$, we have
  \[
  \Service_{\ptime_p}^* \geq 
  S_0\cdot \left(
      1 + \frac{\sqrt{\alpha}-1}{(2k)^{\frac{1}{p-p_0}}}
    \right)^{p-p_0}
    \ \geq\ 
    \frac{S_0}{2k}\cdot
    \alpha^{\frac{p-p_0}{2}}.
  \]
\end{lemma}
\begin{proof}
  By Lemma~\ref{lemma:ServiceLB_GUOMFL}, using $\beta\geq 0$, for all $p>
  p_0$, we get $\ServiceLB{p}\geq \left(1 +
    (\alpha-1)\frac{\gamma_{p-2}}{\gamma_{p-1}}\right)\cdot\ServiceLB{p-1}$.
  Induction on $p$ therefore gives
  \begin{equation}\label{eq:prod_GUOMFL}
      \Service_{\ptime_p}^* \geq \ServiceLB{p} \geq
      \ServiceLB{p_0}\cdot 
      \prod_{i=p_0}^{p-1} \left(1+(\alpha-1)\frac{\gamma_{i-1}}{\gamma_{i}}\right)
  \end{equation}
  for all $p\geq p_0$. Similarly to before, we define
  $\gamma_{\min}=\min\{\gamma_{p_0-1},\dots,\gamma_{p-1}\}$ and
  $\gamma_{\max}=\max\{\gamma_{p_0-1},\dots,\gamma_{p-1}\}$. By
  Lemma~\ref{lemma:gammabound_GUOMFL}, the assumptions regarding $p_0$, and
  because the values $\eta_i$ are non-decreasing in $i$, we have
  \begin{eqnarray*}
    \gamma_{\min} & \geq & \frac{\eta_{p_0-1}}{k} \geq
    \frac{\max\{(\alpha-1)+\beta,2\beta\}}{k}
    \quad\text{and}\\[1mm]
    \gamma_{\max} & \leq & \eta_{p} \leq
    (\alpha-1)\Service_{\ptime_p}^* + \beta \leq
    2(\alpha-1)\Service_{\ptime_p}^*.
  \end{eqnarray*}
  The last inequality follows because
  $\Service_{\ptime_p}^*\geq \ServiceLB{p} \geq \ServiceLB{p_0}\geq
  \max\big\{1, \frac{\beta}{\alpha-1}\big\}$ and by
  applying \eqref{eq:alphabetacondition_GUOMFL}. We can now apply Inequality
  (II) from Lemma~\ref{lemma:fractionsum_GUOMFL} to obtain
  \begin{eqnarray}\nonumber
    \Service_{\ptime_p}^* \geq \ServiceLB{p} & \geq &
    \ServiceLB{p_0} \cdot \left(
      1 + (\alpha-1)\left(\frac{\gamma_{\min}}{\gamma_{\max}}\right)^{\frac{1}{p-p_0}}
    \right)^{p-p_0}\\
    & \geq & \label{eq:implicitbound_GUOMFL}
    \ServiceLB{p_0} \cdot \left(
      1 +
      (\alpha-1)\left(\frac{\max\{(\alpha-1)+\beta,2\beta\}}
        {2k(\alpha-1)\Service_{\ptime_p}^*}\right)^{\frac{1}{p-p_0}}
    \right)^{p-p_0}.
  \end{eqnarray}
  In the following, assume that
  \begin{equation}\label{eq:ServiceUB_GUOMFL}
    \Service_{\ptime_p}^* \leq 
    \max\{1,\frac{\beta}{\alpha-1}\}
    \alpha^{\frac{p-p_0}{2}}.
  \end{equation}
  Note that if \eqref{eq:ServiceUB_GUOMFL} does not hold, the claim of the
  lemma is trivially true. By replacing $\Service_{\ptime_p}^*$ on the
  right-hand side of \eqref{eq:implicitbound_GUOMFL} with the upper bound of
  \eqref{eq:ServiceUB_GUOMFL}, we obtain
  \begin{eqnarray*}
    \Service_{\ptime_p}^* \geq \ServiceLB{p} & \geq &
    \ServiceLB{p_0} \cdot \left(
      1 + (\alpha-1)\cdot \left(
        \frac{(\alpha-1) + \beta}
        {2k(\alpha-1)\max\{1,\frac{\beta}{\alpha-1}\}\alpha^{\frac{p-p_0}{2}}}
      \right)^{\frac{1}{p-p_0}}
    \right)^{p-p_0}\\
    & \geq &
    \ServiceLB{p_0} \cdot \left(
      1 + \frac{\alpha-1}
      {(2k)^{\frac{1}{p-p_0}}\sqrt{\alpha}}
    \right)^{p-p_0}\\
    & \geq &
    \ServiceLB{p_0} \cdot \left(
      1 + \frac{\sqrt{\alpha}-1}{(2k)^{\frac{1}{p-p_0}}}
    \right)^{p-p_0}
    \ \geq\ 
    \frac{\ServiceLB{p_0}}{2k}\cdot
    \alpha^{\frac{p-p_0}{2}}.
  \end{eqnarray*}
  The lemma then follows because we assumed that
  $\ServiceLB{p_0}\geq \max\big\{1,\frac{\beta}{\alpha-1}\big\}$.
\end{proof}
\subsubsection{Optimal Offline Algorithm Total Cost}
\label{sec:offlinetotalcost_GUOMFL}

\para{Service Cost}
In order to minimize the service cost, we can simply bound the service cost of \opt\ \ as follows
\[
	\Service_{\ptime_p}^\opt \geq \Service_{\ptime_p}^*.
\]

\para{Movement Cost}
To simplify our analysis, we take no notice of movement cost
by optimal offline algorithm since it has no substantial effect on
the competitive factor we provide since \opt\ \ has to pay at least the optimal service cost
which we show it is large enough.
The total cost of optimal offline algorithm, therefore, is bounded
as follows
\begin{equation}\label{eq:opttotalcostlowerbound_GUOMFL}
	\cost_{\ptime_p}^\opt= \move_{\ptime_p}^\opt+\Service_{\ptime_p}^\opt
	\geq \Service_{\ptime_p}^*.
\end{equation}
\subsubsection{\DGA\ Total Cost}
\label{sec:onlinetotalcost_GUOMFL}
\para{Service Cost}
The online algorithm \DGA\ has to keep the service
cost smaller than a linear function of optimal service cost
as mentioned in \eqref{eq:algguarantee_GUOMFL}.
Thus
\begin{equation}\label{eq:servicecostUP_GUOMFL}
	\Service_{\ptime_p}^\dga < \fpar \Service_{\ptime_p}^* + \spar.
\end{equation}

\para{Movement Cost}
First, using Definition~\ref{def:phases_GUOMFL} we bound the number
of movement in each phase.
\begin{observation}\label{obs:atmostkmoves_GUOMFL}
  For each Phase $p\geq1$, we have $\lambda_p\leq k$.
\end{observation}
\begin{proof}
  As an immediate consequence of
  Definition~\ref{def:phases_GUOMFL}, we obtain that the maximum number of movements in each
 phase is at most $k$. Let $m>m_p$ and consider the movements $[m_p,m]$.
 We prove that if $m<m_{p+1}$, no two the movements in $[m_p,m]$ move the same
  facility. The claim then follows because there are only $k$
  facilities. For the sake of contradiction, assume that there is some
  facility $i$ that is moved more than once and let $m'$ and $m''$
  ($m',m''\in[m_p,m]$, $m'<m''$) be the first two movements in
  $[m_p,m]$, where facility $i$ is moved. We clearly have
  $\dest_{m'}=\src_{m''}$ and Definition~\ref{def:phases_GUOMFL} thus leads to a
  contradiction to the assumption that $m<m_{p+1}$.
\end{proof}
As a result of above observation and
Lemma~\ref{lemma:betagrowth_GUOMFL} and Lemma~\ref{lemma:alphagrowth_GUOMFL}, it is possible to
prove the following lemma to bound the number of \DGA\ movements
by means of optimal service cost.

\begin{lemma}\label{le:movementcostUP_GUOMFL}
  For any $\alpha \geq 1$ and $\beta$ satisfying
  \eqref{eq:alphabetacondition_GUOMFL}, there is a deterministic online
  algorithm \dga\, such that for all times $t\geq 0$,
  the total movement cost $\move_t^{\dga}$ is bounded as
  follows.
  \begin{itemize}
  \item If $\alpha=1$, for any $\ell\geq1$, $\eps>0$, and $\beta\geq
    k(2k)^{1/\ell}/\eps$, we have 
    \[
    \move_t^{\dga} \leq \eps\cdot\Service_t^* + O(\ell k). 
    \]

  \item For $\alpha\geq 1+\eps$ where $\eps>0$ is some constant and any
  $\beta$ satisfying \eqref{eq:alphabetacondition_GUOMFL}, we have 
    \[
    \move_t^{\dga} \leq k \cdot O\left(1+ \log_{\alpha} \Service_t^* + 
    \min\{\frac{\log k}{\log\log k},\log_{\alpha} k\} +
    \log_{\alpha}\frac{k}{1+\beta}\right).
    \]
  \end{itemize}
\end{lemma}
\begin{proof}
  First note that by Observation~\ref{obs:atmostkmoves_GUOMFL}, the movement
  cost of our algorithm by time $\ptime_p$ is at most
  \begin{equation}\label{eq:phasemovecost_GUOMFL}
    \move_{\ptime_p}\leq (p-1)k+1\leq pk.    
  \end{equation}
  Together with the lower bounds on $\Service_{\ptime_p}^*$ of
  Lemma~\ref{lemma:betagrowth_GUOMFL} and Lemma~\ref{lemma:alphagrowth_GUOMFL}, this allows to
  derive an upper bound on the movement cost of our algorithm as a
  function of $\Service_{\ptime_p}^*$. Note that as all upper bound
  claimed in the lemma have an additive term of $O(k)$ (with no 
  specific constant), it is sufficient to prove that the lemma holds
  for all time $t=\ptime_p$, where $p\geq 2$ is a phase number.

  Let us first consider the case where $\alpha=1$. Because in that
  case $\beta/(\alpha-1)$ is unbounded, we can only apply
  Lemma~\ref{lemma:betagrowth_GUOMFL} to upper bound the movement cost as a
  function of $\Service_t^*$. We choose $\ell\geq 1$ and assume that
  $\beta\geq k(2k)^{1/\ell}/\eps$ for $\eps >0$. Together with \eqref{eq:phasemovecost_GUOMFL}, 
  for $p\geq \ell+2$, Lemma~\ref{lemma:betagrowth_GUOMFL} then gives
  \begin{equation}\label{eq:optcostlowerboundedmovecost_GUOMFL}
  \Service_{\ptime_p}^* \geq
  \frac{k(2k)^{\frac{1}{\ell}}}{\eps}\cdot(p-2)\cdot(2k)^{-\frac{1}{\ell}} 
  = \frac{k}{\eps}(p-2) \geq \frac{1}{\eps}(\move_{\ptime_p} - 
  2k). 
  \end{equation}
  The first part of Lemma~\ref{le:movementcostUP_GUOMFL} then follows because the
  total movement cost for the first $\ell+2$ phases is at most $O(\ell
  k)$. The special cases are obtained as follows. For 
  $\beta=\Omega\left(k+k/\eps\right)$, we set $\ell=\Theta(\log 
  k)$ and every $\eps >0$, whereas for $\beta=\Omega(k\log k/\log \log k)$, we set 
  $\eps=\Theta(\log \log k/\log^{1-\delta}k)$ and $\ell=\Theta\big(\frac{1}{\delta} \cdot \frac{\log k}{\log\log
    k}\big)$ for constant $0 < \delta \leq 1$.

  Let us therefore move to the case where $\alpha>1$. Let $p_0$ be the
  first Phase $p_0\geq 2$ for which $\Service_{\ptime_{p_0}}^*\geq
  S_0$, where
  $S_0=\max\big\{1,\frac{\beta}{\alpha-1}\big\}$ as in
  Lemma~\ref{lemma:alphagrowth_GUOMFL}. Further, we set $p_1=p_0 +
  \lceil2\log_\alpha(2k)\rceil$. Using Lemma~\ref{lemma:alphagrowth_GUOMFL},
  for $p\geq p_1$, we have
  \[
  \Service_{\ptime_p} \geq
  \frac{S_0}{2k}\cdot\alpha^{\frac{p-p_1}{2}}\alpha^{\frac{p_1-p_0}{2}}
  \geq S_0\cdot \alpha^{\frac{p-p_1}{2}}.
  \]
  We therefore get
  \[
  \move_{\ptime_p} \leq k\cdot p \leq k 
  \left(p_1 + 2\log_{\alpha}\frac{\Service_{\ptime_p}^*}{S_0}\right)
  \leq k \left(p_0 + 1 + 
    2\log_{\alpha}\Service_{\ptime_p}^* + \log_{\alpha}\frac{2k}{S_0}
  \right).
  \]
  The second claim of Lemma~\ref{le:movementcostUP_GUOMFL} then follows by showing
  that 
  \[
  p_0=O\big(\min\big\{\frac{\log k}{\log\log k}, \log_{\alpha}
  k\big\}\big).
  \]
  If $S_0=1$, we have $p_0=2$. Otherwise, we
  can apply Lemma~\ref{lemma:betagrowth_GUOMFL} to upper bound $p_0$ as the
  smallest value $p_0$ for which
  $\frac{\beta}{\alpha-1}=\beta(p-2)(2k)^{-1/(p-2)}$. For
  $\alpha=O\big(\frac{\log k}{\log\log k}\big)$, the assumption that
  $\alpha$ is at least $1+\eps$ for some constant $\eps>0$ gives that
  $p_0=\Theta\big(\frac{\log k}{\log\log k}\big)$. Otherwise, (i.e.,
  for large $\alpha$), we obtain $p_0
  =\Theta(\log_{\alpha-1}k)=\Theta(\log_{\alpha} k)$.
\end{proof}
Note that by choosing $\alpha>1$, the dependency of the movement
cost $\move_t^\dga$ on the optimal service cost $\Service_t^*$ is only
logarithmic because terms $\min\{\frac{\log k}{\log\log k},\log_{\alpha} k\}$ and $\log_{\alpha}\frac{k}{1+\beta}$
are dominated by $\log k$.
\begin{proof}[{\bf Proof of Theorem~\ref{thm:UB_GUOMFL}}]
Putting $\eqref{eq:opttotalcostlowerbound_GUOMFL}$, $\eqref{eq:servicecostUP_GUOMFL}$, and
Lemma~\ref{le:movementcostUP_GUOMFL} all together
conclude the claim of theorem.
\end{proof}

\section{OMFL: A Lower Bound Analysis}
\label{sec:OMFL}
In this section, we provide a proof of Theorem~\ref{thm:LB_OMFL}. The lower bound presented here, combined with the upper bound presented in Section~\ref{sec:G-OMFL}, provides a tight analysis for G-OMFL on uniform metrics considering Remark~\ref{rem:lowerbound}. Before delving into the details, we overview the lower bound analysis.

\para{Outline of the Analysis} We consider a metric space with uniform distances, i.e., by scaling
appropriately, we can assume that the distance between each
pair of points is  equal to $1$.
Assume that we are given an online algorithm \on\ which guarantees that
\begin{equation}\label{eq:LB_bounds_OMFL}
	\Service^{\on}_t < \Service_t^* + \spar
\end{equation}
at all times $t$ for some parameter $\beta$.
In the following, let \opt\ be any optimal offline algorithm.
We essentially compute the total cost of \on\ and \opt\ at any time $t$ as functions of
the optimal assignment cost at time $t$. Given \on, we construct an execution in which
\on\ has to perform a large number of movements while the optimal assignment cost does
not grow too much. We divide time into
phases such that in each phase, \on\ has to move $\Omega(k)$ facilities
and the optimal assignment cost grows as slowly as possible. For $p$
phases, we define a sequence of integers $n_1\geq n_2\geq\dots \geq n_p$ where 
$n_1 \leq k/3$ and $n_p \geq 1$ and values
$\Gamma_1<\Gamma_2<\dots< \Gamma_p$. In the following, let $v$ be a
free node if $v$ does not have a facility. Roughly, at the beginning of
a  Phase $i$, we choose a set $N_i$ of $n_i$ (preferably) free nodes and
make sure that all these nodes have $\Gamma_i$ requests. Note that
constructing an execution means to determine where to add the request
in each iteration. The value $\Gamma_i$ is chosen large enough such
that throughout  Phase $i$ a assignment cost of $n_i\Gamma_i$ is
sufficiently large to force an algorithm to move. Hence, whenever
there are $n_i$ free nodes with $\Gamma_i$ requests, \on\ has to move
at least one facility to one of these nodes. For each such movement, we
pick another free node that currently has less than $\Gamma_i$
requests and make sure it has $\Gamma_i$ requests. We proceed until
there are $k$ nodes with $\Gamma_i$ requests at which point the main
part of the phase ends. Except for the nodes in $N_i$, each of the $k$
nodes with $\Gamma_i$ requests leads to a movement of \on\ and
therefore, \on\ has to move at least $k-n_i=\Omega(k)$ facilities in
 Phase $i$.  At the end of  Phase $i$, we can guarantee that there are
exactly $k$ nodes with $\Gamma_i$ requests, $n_i$ nodes with
$\Gamma_{i-1}$ requests, $n_{i-1}-n_i$ nodes with $\Gamma_{i-2}$
requests, etc. The optimal assignment cost after  Phase $p$, therefore, is
$n_p\Gamma_{p-1}+\sum_{i=3}^{p} (n_{i-1}-n_i)\Gamma_{i-2}$.  The
assignment cost paid by \on \ at time $t$ can not be smaller than
$\Service^*_t$.  

By contrast, the optimal offline algorithm can wait until all requests have arrived and
just perform all the necessary facility movements at the very end to have an optimal
configuration. Therefore by the end of  Phase $p$, \opt\ has to pay at most $k$ as the
total movement cost, while \on\ has to pay $\Theta(pk)$ for the movement cost in total by this time.
The assignment cost of \opt \ equals the optimal
assignment cost at the end of  Phase $p$. By choosing the values $n_i$
appropriately, we obtain the claimed bounds.

\subsection{Lower Bound Analysis}
\label{sec:lowerbound_GUOMFL}
The formal proof consists of three
parts. In Section~\ref{sec:LBexecution_OMFL}, given some online algorithm
\on, we construct an explicit bad execution. In
Section~\ref{sec:loweralgtotalcost}, we analyze the cost of the online
algorithm \on\ in the constructed execution and in
Section~\ref{sec:lowerofflinealftotalcost_GUOMFL}, we bound the cost of an
optimal offline algorithm \opt\, and we combine everything to
complete the proof of Theorem~\ref{thm:LB_OMFL}.

\subsubsection{Lower Bound Execution}
\label{sec:LBexecution_OMFL}
 We assume that \on\ is the given
online algorithm and \opt\ is an optimal offline
algorithm. Further recall that we assume that \on\ guarantees that
the difference between the assignment cost of \on\ and the optimal
assignment cost at all times is less than $\beta$ for some given
$\beta>0$.

  We need $n$ to be sufficiently large and for simplicity,
we assume that $n\geq 3k$. We denote a feasible configuration by a set
$F\subset V$ of size $\lvert F\rvert=k$. Further, without loss of generality, we
assume that all facilities of \on\ and \opt\ are at the same
locations at the beginning (i.e. at time $t=0$). At each point $t$ in
the execution, a configuration $F_t^*$ with optimal assignment cost
places facilities at the $k$ nodes with the most requests (breaking ties
arbitrarily if there are several nodes with the same number of
requests). Also, at a time $t$ the optimal assignment cost is equal to
the total number of requests at nodes in $V\setminus F_t^*$ for an
arbitrary optimal configuration $F_t^*$.

  Time is divided into phases. We construct the execution such that
  it lasts for at least $k$ phases. As described in the
  outline, we define integers $\Gamma_1<\Gamma_2<\dots$ such that at
  the end of  Phase $i$, there are exactly $k$ nodes with $\Gamma_i$
  requests (and all other nodes have fewer requests). For each phase
  $i$, we define $V_i$ to be this set of $k$ nodes with $\Gamma_i$ requests. We also fix
  integers $n_1\geq n_2\geq \dots\geq 1$ where $n_1 \leq k/3$ and at the beginning
  of each  Phase $i$, we pick a set $N_i$ of $n_i$ nodes to which we
  directly add requests so that all of them have exactly $\Gamma_i$
  requests. For $i=1$, we pick $N_1$ as an arbitrary subset of
  $V\setminus F_0$. We define $V_0:=F_0$. For $i\geq2$, we choose $N_i$ as an arbitrary subset
  of $V_{i-2}\setminus V_{i-1}$. Clearly, at the end of  Phase $i$, we
  have $N_i\subseteq V_i$ as otherwise there would be more than $k$
  nodes with exactly $\Gamma_i$ requests. Note that because
  $N_{i-1}\subseteq V_{i-1}$ and because $N_{i-1}\cap
  V_{i-2}=\emptyset$, $V_{i-2}\setminus V_{i-1}$ contains $n_{i-1}\geq
  n_i$ nodes and it is therefore possible to choose $N_i$ as
  described. Note also that because $N_i\subseteq V_{i-2}\setminus
  V_{i-1}$, at the beginning of  Phase $i$ all nodes in $N_i$ have
  exactly $\Gamma_{i-2}$ requests. The remaining ones of the $k$ nodes
  that end up in $V_i$ (and thus have $\Gamma_i$ requests at the end of
   Phase $i$) are chosen among the nodes in $V_{i-1}$. Consequently, at
  the end of  Phase $i-1$ and thus at the beginning of  Phase $i$, there
  are exactly $k$ nodes $V_{i-1}$ with $\Gamma_{i-1}$ requests,
  $n_{i-1}$ nodes $V_{i-2}\setminus V_{i-1}$ with $\Gamma_{i-2}$
  requests, $n_{i-2}-n_{i-1}$ requests $V_{i-3}\setminus(V_{i-2}\cup
  N_{i-1})$ with $\Gamma_{i-3}$ requests, $n_{i-3}-n_{i-2}$ nodes with
  $\Gamma_{i-4}$ requests, and so on.
  Now, $n_i$ of the nodes in
  $V_{i-2}\setminus V_{i-1}$ are chosen as set $N_i$ and we increase
  their number of requests to $\Gamma_i$. From now on, throughout phase
  $i$, there are $k+n_i$ nodes with at least $\Gamma_{i-1}$ requests
  such that at most $k$ of these nodes have $\Gamma_i$ requests. The
  number of nodes with less than $\Gamma_{i-1}$ requests is the same as
  at the end of  Phase $i-1$. In fact nodes that are not in
  $V_{i-1}\cup N_i$ do not change their number of requests after phase
  $i-1$. As a consequence of the execution,
  after increasing the number of requests in
  $N_i$ to $\Gamma_i$, the optimal assignment cost remains constant
  throughout  Phase $i\geq 1$ and it can be evaluated to
  \[
  \Sigma_i^* := n_i \cdot \Gamma_{i-1} + \sum_{j=2}^{i-1} (n_j - n_{j+1})\Gamma_{j-1}.
  \]
  For convenience, we also define $\Sigma_0^*:=0$ and moreover $\Sigma_1^*=0$ since
  there are at most $k$ nodes with $\Gamma_1$ requests at the end of  Phase $1$.

    In the following, let $v$ be a free node at some point in the
  execution, if the algorithm currently has no facility at node
  $v$. We now fix a  Phase $p\geq 1$ and assume that we are at a time
  $t$, when we have already picked the set $N_p$ and increased the
  number of requests of nodes in $N_p$ to $\Gamma_p$. By the above
  observation, we have $\Service_t^*=\Sigma_p^*$ and therefore \on\
  is forced to move if there are $n_p$ free nodes with $\Gamma_p$
  requests and if we choose $\Gamma_p$ such that
  \begin{equation}\label{eq:LBGamma_OMFL}
    \gamma_p := \Gamma_p-\Gamma_{p-1} = \frac{(\alpha-1)\Sigma_p^*+\beta}{n_p}.
  \end{equation}
  We can now describe how and when the remaining $k-n_p$ nodes of
  $V_p$ are chosen after picking the nodes in $N_p$. As described
  above, the nodes are chosen from $V_{p-1}$. We
  choose the nodes sequentially. Whenever we choose a new node from
  $V_p$, we pick some free node $v\in V_{p-1}$ with less than
  $\Gamma_p$ requests and increase the number of requests of $v$ to
  $\Gamma_p$. As described above, $\Gamma_p$ is chosen large enough
  (as given in \eqref{eq:LBGamma_OMFL}) such that throughout  Phase $p$
  there are never more than $n_p-1$ free nodes with $\Gamma_p$
  requests. Because $\lvert N_p\cup V_{p-1}\rvert=k+n_p$, as long as there are at
  most $k$ nodes with $\Gamma_p$ requests there always needs to be a
  free node $v\in V_{p-1}$ that we can pick and we actually manage to
  add $k$ nodes to $V_p$.

\subsubsection{Online Algorithm Total Cost}
\label{sec:loweralgtotalcost}
  The assignment cost paid by \on \ at any time $t$ could be simply lower bounded
  by $\Service^*_t$. Hence, it remains to compute a lower bound for 	$\move_t^\on$ as a function
of optimal assignment cost. The following lemma computes such a lower bound.
  
  \begin{lemma}\label{le:onlinealgmoveboundLB_GUOMFL}
  For any $\alpha \geq 1$ and $\beta$, assume \on \ be any deterministic online
  algorithm that can solve the problem. There exists a time $t > 0$ such that the execution
  of Section~\ref{sec:LBexecution_OMFL} guarantees
  the total movement cost $\move_t^\on$ can be
  bounded as follows.
 
  \begin{itemize}
  \item If $\alpha=1$, for any $\ell\geq1$, $\eps>0$, and $\beta\leq
    k(2k)^{1/\ell}/\eps$, we have 
    \[
    \move_t^{\on} \geq \eps\cdot\Service_t^* + \Omega(\ell k). 
    \]
    
    Specifically, for $\beta=O(k/\eps)$ we get
    $\move_t^{\on} \geq \eps \cdot \Service_t^* + \Omega(k\log k)$ and for
    $\beta=O\big(\frac{k\log k}{\log\log k}\big)$ we have $\move_t^{\on} \geq \eps \cdot \Service_t^* +
    \Omega\big(\frac{k\log k}{\log\log k}\big)$.  \medskip

  \item For $\alpha\geq 1+\eps$ where $\eps>0$ is some constant and any $\beta$, we have 
    \[
    \move_t^{\on} \geq k \cdot \Omega\left(1+ \log_{\alpha} \Service_t^*\right).
    \]
  \end{itemize}
\end{lemma}
\begin{proof}
Let us count the number of movements of \on\ in a given Phase
  $p$. 
  At each point in time $t$ during the phase, let $\Phi_t$ be the
  number of free nodes with $\Gamma_p$ requests (possibly including a
  node $v$ that we already chose to be added to $V_p$). We know that
  for all $t$, $\Phi_t<n_p$. Whenever we decide to add a new node $v$
  to $V_p$, $\Phi_t$ increases by $1$ (as $v$ is a free node). The
  value of $\Phi_t$ can only decrease when \on\ moves a facility and
  each facility movement reduces the value of $\Phi_t$ by at most
  $1$. As after fixing $N_p$, we add $k-n_p$ nodes to $V_p$, we need
  at least $k-2n_p\geq k/3$ movements to keep $\Phi_t$ below $n_p$
  throughout the phase. Consequently, every online algorithm \on\ has
  to do at least $k/3$ movements in each phase.

  Now we upper bound the optimal assignment
  cost $\Sigma_p^*$ as a function of $\alpha$, $\beta$, and $p$. Using
  \eqref{eq:LBGamma_OMFL}, for all $p\geq 0$, we have
  \[
  \Sigma_p^* = \sum_{i=1}^p n_p\cdot \gamma_{p-1} 
  \]
  For $p\geq 1$, we then get
  \begin{equation}\label{eq:ServiceLower_GUOMFL}
  \begin{aligned}
    \Sigma_p^* & =
    \frac{n_p}{n_{p-1}}\big((\alpha-1)\Sigma_{p-1}^*+\beta\big)
    +\Sigma_{p-1}^*\\
    & = \left(1 + (\alpha-1)\frac{n_p}{n_{p-1}}\right)\cdot \Sigma_{p-1}^*
    + \beta\cdot \frac{n_p}{n_{p-1}}.
  \end{aligned}
  \end{equation}
  In the following, we for simplicity assume that for $i=1,2,\dots,p$,
  values $n_i$ do not have to be integers. For integer $n_i$, the proof
  works in the same way, but becomes more technical and harder to
  read. We fix the values of $n_i$ as 
  \[
  n_i:=(k/3)^{\frac{p-i}{p-1}}
  \]
  such that $n_1=k/3$ and $n_p=1$. For all $i\geq 1$, we then have
  $\frac{n_i}{n_i-1}=(\frac{k}{3})^{-\frac{1}{p-1}}$.
  \eqref{eq:ServiceLower_GUOMFL} now be simplified as
  \begin{equation}\label{eq:mainLB_GUOMFL}
  \Sigma_p^* = \left(1 +
    \frac{\alpha-1}{(k/3)^{1/(p-1)}}\right)\cdot\Sigma_{p-1}^*+ \beta\cdot \frac{1}{(k/3)^{1/(p-1)}}.
  \end{equation}
  We have already seen that $\Service_t^*=\Sigma_p^*$.
  Using \eqref{eq:mainLB_GUOMFL} and \eqref{eq:ServiceLower_GUOMFL}, the claim of the first part of the lemma follows
  analogously from Lemma~\ref{lemma:ServiceLB_GUOMFL}
  and Lemma~\ref{lemma:betagrowth_GUOMFL} and the claim of the second part of the lemma follows
  analogously from Lemma~\ref{lemma:ServiceLB_GUOMFL} and Lemma~\ref{lemma:alphagrowth_GUOMFL}
  in the upper bound analysis section. 
  \end{proof}

\subsubsection{Optimal Offline Algorithm Total Cost}
\label{sec:lowerofflinealftotalcost_GUOMFL}

An optimal offline algorithm, say \opt, knows the request sequence
in advance. In other words, it can wait until all requests have arrived and
just perform all the necessary facility movements at the very end. Therefore,
an upper bound for the total cost of \opt\ at any time $t$ is 
\begin{equation}\label{eq:LBopttotalcost_GUOMFL}
	\cost^{\opt}_t \leq k+\Service^*_t.
\end{equation}

We now have everything we need to prove Theorem~\ref{thm:LB_OMFL}.

\begin{proof}[{\bf Proof of Theorem~\ref{thm:LB_OMFL}}]
  The proof of Theorem~\ref{thm:LB_OMFL} now directly follows from
  Lemma~\ref{le:onlinealgmoveboundLB_GUOMFL} and from \eqref{eq:LBopttotalcost_GUOMFL}.
\end{proof}

\section{M-OMFL: A Lower Bound Analysis}
\label{sec:M-OMFL}
We provide our lower bound execution in the following.
We consider a uniform metric where the distance between every pair of points is 1. In this setting, the goal of any feasible solution for M-OMFL is to minimize the total number of movements.
As we can assume that each node either has $0$ or $1$
 facilities, we slightly overload our notation and simply denote a feasible
 configuration by a set $F\subset V$ of size $\lvert F\rvert=k$.
 We first fix \on \ to be any given deterministic online algorithm and \opt\
to be an optimal offline algorithm denoted by \opt. For proving the statement of
Theorem~\ref{thm:LB_MMUOMFL}, we distinguish two cases, depending on
the number of facilities $k$. In both cases, we define
\textit{iterations} to be subsequences of requests such that \on\
needs to move at least once per iteration. The number of movements by
\on\ is therefore at least the number of iterations of a given
execution.
  
\para{Case \boldmath$k \leq \floor{n/2}$} At the beginning, we
place a large number of requests on any $k-1$ nodes that initially
have facilities. We choose this number of requests sufficiently large
such that no algorithm can even move any of these $k-1$ facilities.  This
essentially reduces the problem to $k=1$ and $n-k+1$ nodes.

To bound the number of movements by \opt, we then consider intervals
of $n-k$ iterations such that \on\ is forced to move in each
iteration. During each interval, the requests are distributed in such
a way that at the beginning of the $i$-th iteration of the interval
there are at least $n-k-i+1$ nodes such that if any offline algorithm
places a facility on one of these nodes, \eqref{eq:bounds_MMUOMFL}
remains satisfied throughout the whole interval.  Hence, there exists
an offline algorithm that moves at most once in each interval and
therefore the number of movements by \opt \ is upper bounded by the
number of intervals.
 
\para{Case \boldmath$k > \floor{n/2}$} In this case, there is
some resemblance between the constructed execution and the lower bound
constructions for the paging problem. For simplicity assume that there
are $n=k+1$ nodes (we let requests arrive at only $k+1$ nodes).  At
the beginning of each iteration we locate a sufficiently large number
of requests on the node without any facility of \on\ such that
\eqref{eq:bounds_MMUOMFL} is violated. Thus, \on\ has to move at least one
server to keep \eqref{eq:bounds_MMUOMFL} satisfied. By contrast, \opt\ does
not need to move in each iteration. There is always a node which will
not get new requests for the next $k$ iterations and therefore \opt\
only needs to move at most once every $k$ iterations to keep \eqref{eq:bounds_MMUOMFL}
satisfied. 
  
 \begin{proof}[{\bf Proof of Theorem~\ref{thm:LB_MMUOMFL}:}]
Consider any request sequence. First we provide a partitioning of the request sequence
as follows.  The request sequence is partitioned into \textit{iterations}. Iteration $0$ is the empty
   sequence and for every $i \geq 1$, iteration $i$ consists of a
   request sequence of a length dependent on $\alpha$, $\beta$, and
   the iteration number $i$. The request sequence of an iteration $i$ is
   chosen dependent on a given online algorithm \on\ such that \on\
   must move at least once in iteration $i$. We see that while
   \on\ needs to move at least once per iteration, there is an
   offline algorithm which only moves once every at least $n/2$
   iterations.
   
   In the proof, we reduce all the cases to two extreme cases. In the first case, we reduce
   the original metric on a set of $n$ nodes with $k \leq \floor{n/2}$
   facilities to the case where there is only $1$ facility. To do this, we first
   place sufficiently many requests on $k-1$
   nodes that have facilities at the beginning of execution (for
   simplicity, assume that we place an unbounded number of requests on
   these nodes). This
   prevents any algorithm from moving its facilities from these $k-1$
   nodes during the execution and hence we can ignore these $k-1$
   nodes and facilities in our analysis. In
   contrast, for the second case where $k > \floor{n/2}$, we assume that
   w.l.o.g., $k=n-1$ by simply only placing requests on the $k$ nodes
   which have facilities at the beginning and on one additional node. 

   In the following, we let $t_i$ denote the end of an iteration
   $i$. Moreover suppose $\mathcal{I}$ is the total number of
   iterations, where we assume that
   $\mathcal{I} \equiv 0\pmod{\max\{k,n-k\}}$.
   
   \para{Case \boldmath$k \leq \floor{n/2}$} 
      The idea behind the execution is to uniformly increase the
      number of requests on the $n-k$ nodes that do not have the
      facility at the beginning of an
      iteration $i$ (i.e., at time $t_{i-1}$) in such a way that \on \ has to move at least
      once to satisfy \eqref{eq:bounds_MMUOMFL} at the end of iteration $i$.
      Moreover the distribution of requests guarantees that any node
      without the facility at time $t_{i-1}$ is a candidate
      to have the (free) facility of \on\ at time $t_i$. Let $v$ be any node in the set $V$ of nodes. We use $\ndem_{v,t}$ to denote the number of requests at node $v$ after the arrival of $t$ requests. When the context is clear, we omit the second subscript (i.e., $t$) and simply write $\ndem_v$. Further, let $v^\on_t$
      denote the node on which \on \ locates its facility at time
      $t$ and let $U(t)$ be the set of all nodes without facility at
      time $t$. Moreover, let $v^*_t$ be a
      node which has the largest number of requests among all nodes at time $t$. The node with the largest number of
      requests at the end of an iteration $i$, i.e. $v^*_{t_i}$, is chosen such that $v^*_{t_i} \neq v^\on_{t_{i-1}}$.
      At time $0$, we have $r_u=0$ for all nodes $u$. The distribution of requests at the end of iteration $i$ is
      as follows:  
      \begin{eqnarray}\label{eq:minmove_LB_FC_distreq}
        \forall u \in U(t_{i-1}) \setminus \{v^*_{t_i}\} : r_u &= &
	   r_{v^*_{t_{i-1}}}+\max\{\beta,1\},\\
        \label{eq:minmove_LB_FC_algdistreq}
         r_{v^\on_{t_{i-1}}}&=&r_{v^*_{t_{i-1}}},\\
        \label{eq:minmove_LB_FC_optdistreq}
        r_{v^*_{t_i}}&=& (\alpha-1)
	   \cdot \Service^*_{t_i} + r_{v^*_{t_{i-1}}} + \beta.
      \end{eqnarray}
      \begin{adjustwidth}{4mm}{4mm}
          \begin{claim}\label{cl:minmove_LB_FC}
            The above execution guarantees that \on\ has to move at
            least once per iteration. Further, there exists an offline
            algorithm \off\ that moves its facilities at most
            $\mathcal{I}/(n-k)$ times.
          \end{claim}
          \begin{proof}
            Consider any interval of $n-k$ iterations such that the
            first iteration of this interval has ending time $\tau_1$
            and the finishing time of the last iteration (or the
            finishing time of the interval) is $\tau_{n-k}$. Further,
            suppose the previous interval has finished at $\hat{t}$.
            Obviously, if this is the first interval, $\hat{t}=0$.
            Let
            $U:=U(\hat{t}) \setminus
            \bigcup_{t=\tau_1}^{\tau_{n-k}}\{v^\on_t\}$
            denote the set of nodes which have not had the facility of
            \on \ during this interval. The offline algorithm for all
            iterations of this interval, locates its facility either on
            node $v^\on_{\tau_{n-k}}$ if set $U$ is empty or on
            some node in $U$, otherwise.  The case in which $U$ is
            empty indicates that every node in $U(\hat{t})$ has had
            the facility of \on \ exactly once within the
            interval. Whenever the offline algorithm needs to move, it
            locates its facility at a node in
            $U \cup \{v^\on_{\tau_{n-k}}\}$.  On the one hand and
            according to \eqref{eq:minmove_LB_FC_distreq}, node
            $v^\on_{\tau_{n-k}}$ or any node in $U$ (in the case this
            set is not empty) has at least
            $r_{v^*_{t_{i-1}}}+\max\{\beta,1\}$ requests at the end
            of each iteration $i$ that is in this interval.
            Therefore, the offline assignment cost at $t_i$ is
            \begin{equation}\label{eq:minmove_LB_FC_OFF_SCost}
              \Service^\off_{t_i} \leq (\alpha-1)\cdot \Service^*_{t_i}+2r_{v^*_{t_{i-1}}}+\beta+(n-k-2)
              \cdot \left(\max\{\beta,1\}+r_{v^*_{t_{i-1}}}\right)
            \end{equation}  
            On the other hand, the optimal assignment cost is
            \begin{equation}\label{eq:minmove_LB_FC_OPT_SCost}
              \Service^*_{t_i} = (n-k-1) \cdot \left(\max\{\beta,1\}+r_{v^*_{t_{i-1}}}\right)+ r_{v^*_{t_{i-1}}}
            \end{equation}
            using \eqref{eq:minmove_LB_FC_distreq}, \eqref{eq:minmove_LB_FC_algdistreq}, and \eqref{eq:minmove_LB_FC_optdistreq}.
            Hence
            \eqref{eq:minmove_LB_FC_OFF_SCost} and \eqref{eq:minmove_LB_FC_OPT_SCost} imply that
            \begin{equation}\label{eq:minmove_LB_FC_OFF_SCost_bounded}
              \Service^\off_{t_i} < \alpha \Service^*_{t_i} + \beta.
            \end{equation}
            This guarantees that offline algorithm does not need to
            move more than once during any interval of $n-k$ iterations.
            In other words, at the beginning of the interval, the offline algorithm decides to locate
            its facility to a node in $U \cup \left\{v^\on_{\tau_{n-k}}\right\}$
            if it needs because it knows the behavior of the online algorithm in advance as well as
            the request sequence. According to \eqref{eq:minmove_LB_FC_OFF_SCost_bounded},
            this one movement by the offline algorithm is sufficient to keep \eqref{eq:bounds_MMUOMFL} satisfied
            within the interval. Therefore, the offline
            algorithm moves at most $\mathcal{I}/(n-k)$ times.
            
            At the end of each iteration $i$, if the online algorithm has not moved yet within the iteration $i$
            then we have $v^\on_{t_{i-1}} = v^\on_{t_i}$. Thus,  
            \begin{equation}\label{eq:minmove_LB_FC_ALG_SCost}
    	      \Service^\on_{t_i} = (\alpha-1)\cdot \Service^*_{t_i}+r_{v^*_{t_{i-1}}}+\beta+(n-k-1)
	      \cdot \left(\max\{\beta,1\}+r_{v^*_{t_{i-1}}}\right)
            \end{equation}  
            with respect to \eqref{eq:minmove_LB_FC_distreq}, \eqref{eq:minmove_LB_FC_algdistreq}, and
            \eqref{eq:minmove_LB_FC_optdistreq}. Therefore due to \eqref{eq:minmove_LB_FC_OPT_SCost}
            and \eqref{eq:minmove_LB_FC_ALG_SCost}
            we have $\Service^\on_{t_i} = \alpha\Service^*_{t_i} +\beta$.
            This implies that the online algorithm must had moved at least once to guarantee 
            \[
            \forall i : v^\on_{t_{i-1}} \neq v^\on_{t_i}.
            \] 
            Thus \on \ has to move once per iteration and then the claim holds.
          \end{proof}
          \begin{corollary}\label{co:minmove_LB_FC}
            The Claim~\ref{cl:minmove_LB_FC} implies that 
            \[
            \move_t^\off \leq \frac{\move_t^\on}{n-k}.
            \]
            where $t$ be the ending time of $(c \cdot (n-k))$-th iteration for any integer $c \geq 1$.
          \end{corollary}
          \begin{proof}
            It follows by the fact that $\move_t^\opt \leq \move_t^\off$.
          \end{proof}
      \end{adjustwidth}

    \para{Case \boldmath$k > \floor{n/2}$}
       Here when we have more facilities than half of the nodes, we assume, w.l.o.g. $n=k+1$.
       This is doable by letting the requests arrive at a fix set of nodes of size $k+1$ including $k$ facilities. 
       Therefore, at each time
       there is only one node without a facility in which this situation holds for any algorithm.
       Let $\bar{v}^\on_t$ denote
       the node without any facility of \on \ at time $t$. We force \on \ to move in each iteration
       $i$ by putting large enough number of requests
       on $\bar{v}^\on_{t_{i-1}}$ while any optimal offline algorithm only moves one of its facilities after
       at least $k$ iterations. Consider
       an interval of $k$ iterations starting from the first iteration of this interval with ending time
       $\tau_1$ and ending at the last iteration
        at time $\tau_{k}$. For any iteration $i$ of this interval the distribution of the requests at
        the end of the iteration is as follows.
       \begin{equation}\label{eq:minmove_LB_SC_distreq}
	    r_{\bar{v}^\on_{t_{i-1}}}=\alpha \Service^*_{t_i} + \max\{\beta,1\}.
       \end{equation}  
       According to \eqref{eq:minmove_LB_SC_distreq} the optimal assignment cost does not change
       during the interval, i.e.
       $\Service^*_{\tau_i}=\Service^*_{\tau_{i}+1}$ for all $i \in
       [k-1]$ of the current interval.
       \begin{adjustwidth}{4mm}{4mm}
           \begin{claim}\label{cl:minmove_LB_SC}
             The above execution guarantees that \on \ has to move at least once per iteration
             while the number of movements      
             by any optimal offline algorithm is at most $\mathcal{I}/k$.
           \end{claim}
           \begin{proof}
             At the end of iteration $i$, assume $\bar{v}^\on_{t_{i-1}} = \bar{v}^\on_{t_i}$, then we have
             \begin{equation}\label{eq:minmove_LB_SS_ALG_SCost}
               \Service^\on_{t_i}=\alpha \Service^*_{t_i} + \max\{\beta,1\} \geq \alpha\Service^*_{t_i} + \beta
             \end{equation}      
             using \eqref{eq:minmove_LB_SC_distreq}. It implies that the online algorithm
             must had moved at least once to guarantee
             \[
             \forall i : \bar{v}^\on_{t_{i-1}} \neq \bar{v}^\on_{t_i}.
             \]

             The optimal offline algorithm, by contrast, need to move a facility from $\bar{v}^\on_{\tau_k}$
             to $\bar{v}^\on_{\tau_1}$ during
             the interval with respect to the request distribution in \eqref{eq:minmove_LB_SC_distreq}.
             The node $\bar{v}^\on_{\tau_k}$ is the node has $\alpha \Service^*_{\hat{t}} + \max\{\beta,1\}$ requests
             within the interval due to \eqref{eq:minmove_LB_SC_distreq} where $\hat{t}$ is the ending
             time of any iteration of the previous interval.
             Hence, at the end of any iteration $i$ in the interval, the optimal offline assignment cost equals
             the optimal assignment cost and thus
             \eqref{eq:bounds_MMUOMFL} remains satisfied. Consequently it implies that at most one
             movement by optimal offline algorithm is sufficient during
             the interval. This concludes that the number of movements by any optimal offline algorithm
             is at most $\mathcal{I}/k$ in this case. 
           \end{proof}
       \end{adjustwidth}
      Let $t$ be the ending time of $(c \cdot \max\{k,n-k\})$-th iteration for any integer $c \geq 1$.
      Using Corollary~\ref{co:minmove_LB_FC} and Claim~\ref{cl:minmove_LB_SC}
      \[
 	   \move_t^\on \geq \max\{n-k,k\} \cdot \move_t^\opt \geq \frac{n}{2} \cdot \move_t^\opt.
      \]	
      Thus the claim of the theorem holds.
\end{proof}

\section{Conclusion}
\label{sec:conclusion}
In light of the limited research on the online variants of the mobile facility location problem (MFL), we introduce and examine the OMFL problem and its two subtypes: G-OMFL and M-OMFL. We establish tight bounds for G-OMFL on uniform metrics, where our lower bound for OMFL also applies to G-OMFL on uniform metrics since OMFL on uniform metrics is a special case of G-OMFL on uniform metrics. Additionally, we demonstrate a linear lower bound on the competitiveness for M-OMFL, even on uniform metrics.

Motivated by the approach used by \cite{cote2008randomized,bansal2011polylogarithmic} for the $k$-server problem, we define and study G-OMFL on uniform metrics, similar to the allocation problem defined by \cite{cote2008randomized,bansal2011polylogarithmic} on uniform metrics. This is the first step towards solving OMFL on HSTs and general metrics. The second step, which remains an open question, is to adapt a similar approach for OMFL using the \DGA\ algorithm presented in this paper. The idea is for each internal node of the HST to run an instance of G-OMFL to decide how to allocate its facilities among its children nodes. Starting from the root, which has $k$ facilities, this recursive process determines the number of facilities at each leaf of the HST, providing a feasible solution for OMFL.

The second open question is whether the result provided by Theorem~\ref{thm:LB_MMUOMFL} is tight. Additionally, due to the similarities between the M-OMFL problem in uniform metrics and the paging problem, it would be interesting to explore the use of randomized online algorithms against oblivious adversaries for M-OMFL. The goal would be to achieve a sublinear competitive ratio for this problem.

\section*{Acknowledgments}
Abdolhamid Ghodselahi would like to express his gratitude for the support received from the joint research project DIG-IT!, which is funded by the European Social Fund (ESF), reference: ESF/14-BM-A55-0017/19, and the Ministry of Education, Science and Culture of Mecklenburg-Vorpommern, Germany. Moreover, the majority of this work was done while Abdolhamid Ghodselahi was at the Department of Computer Science at the University of Freiburg, Freiburg, Germany.

\bibliographystyle{abbrv}
\bibliography{references}

\newpage
\appendix

\section{HSTs}\label{sec:app_hst}

Embeddings of a metric space into (a probability distribution over)
tree metrics has found many important applications
\cite{ghods2017arrow,azar2017online,bansal2011polylogarithmic,cote2008randomized}.
The notion of a \emph{hierarchically well-separated tree} (in the following referred to as an HST)
was defined by Bartal in \cite{bartal96}. 

\begin{definition}\label{def:hst}
Given a parameter $\alpha>1$, an $\alpha$-HST of depth $h$ is a rooted tree
with the following properties. All children of the root are at distance $\alpha^{h-1}$
from the root. Further, every subtree of the root is an $\alpha$-HST of depth $h-1$
that is characterized by the same parameter $\alpha$ (i.e., the children
2 hops away from the root are at distance $\alpha^{h-2}$ from their parents). 
A tree is an HST if it is an $\alpha$-HST for some $\alpha>1$.
\end{definition}

The probabilistic tree embedding result of \cite{fakcharoenphol2003tight} shows that for every metric space
$(X,d)$ with minimum distance normalized to $1$ and for every constant $\alpha> 1$, there is a
randomized construction of an $\alpha$-HST $T$ with a bijection $f$ of the points in $X$ to the leaves
of $T$ such that a) the distances on $T$ are dominating the distances
in the metric space $(X,d)$, i.e., $\forall x,y\in X:
d_T(f(x),f(y))\geq d(x,y)$ and such that b) the expected tree distance
is $\E\big[d_T(f(x),f(y))\big] = O(\alpha \log \lvert X \rvert/\log \alpha)\cdot d(x,y)$ for every
$x,y\in X$.

Utilizing HSTs as a tool in solving some online problems is
a common approach
\cite{ghods2017arrow,azar2017online,bansal2011polylogarithmic,cote2008randomized}.
The procedure of finding an approximate optimal solution for a given
online minimization problem is roughly as follows: an HST is sampled
according to the distribution defined by the embedding. The problem is
then solved on the HST with a competitive ratio of $\gamma$. Since the
distances in the HST are at least the corresponding distances in the original metric, the solution on the HST provides a solution on
the original graph of at most the same cost.  However, to bound the
cost of the optimal solution w.r.t. the distances in the HST
from above by the corresponding distances in the original metric, we
lose a $O(\log n)$ factor. Consequently, we get an expected
$O(\gamma \cdot \log n)$ competitive ratio.

\end{document}